\documentclass[journal]{IEEEtran}

\usepackage{tikz}
\usepackage{pgfplots}
\pgfplotsset{compat=1.17}
\usetikzlibrary{external}
\usetikzlibrary{patterns}
\usepackage{array}
\usepackage{lipsum,graphicx,multicol,multirow}
\usetikzlibrary{automata,chains,shadows,fit,backgrounds,shapes,arrows,positioning,calc,decorations.pathreplacing}
\usetikzlibrary{matrix}
\usetikzlibrary {arrows.meta,shapes.misc,calc}
\usetikzlibrary{graphs,graphs.standard,quotes,fillbetween}
\usepackage{amsmath, amssymb,amsfonts,amscd,xpatch}
\usepackage{cases}
\usepackage{mwe,comment,balance}
\usepackage{xcolor,colortbl} 
\usepackage{pgfplotstable}
\usepgfplotslibrary{fillbetween} 
\usepackage[noadjust]{cite}

\usepackage{arydshln}

\usepackage[linesnumbered,ruled]{algorithm2e}

\usepackage{booktabs}

\usepackage{url}

\newcommand{\Var}[1]{\mathrm{Var}\left[#1\right]}

\newcommand{\argmin}{\mathop{\mathrm{argmin}}}

\newcommand{\yeq}{\overline{y}}
\newcommand{\byeq}{\overline{\boldsymbol{y}}}

\newcommand{\xhat}{\hat{x}}

\newcommand{\PsVec}[1]{\boldsymbol{P}_{#1}}

\newcommand{\lnalpha}{\tilde{\alpha}}
\newcommand{\lnbeta}{\tilde{\beta}}
\newcommand{\lngamma}{\tilde{\gamma}}
\newcommand{\lnxi}{\tilde{\xi}}

\newcommand{\bx}{\boldsymbol{x}}
\newcommand{\bxhat}{\hat{\boldsymbol{x}}}

\newcommand{\by}{\boldsymbol{y}}

\newcommand{\mcX}{\mathcal{X}}
\newcommand{\mcS}{\mathcal{S}}
\newcommand{\mcN}{\mathcal{N}}

\newcommand{\nbA}{\boldsymbol{A}}

\newcounter{lemma}

\newtheorem{example}{Example}

\ifCLASSOPTIONcompsoc
 \usepackage[caption=false,font=normalsize,labelfont=sf,textfont=sf]{subfig}
\else
 \usepackage[caption=false,font=footnotesize]{subfig}
\fi

\newcommand{\NoRev}[1]{{\color{black}#1}}
\newcommand{\RevA}[1]{{\color{black}#1}}
\newcommand{\RevB}[1]{{\color{black}#1}}
\newcommand{\RevC}[1]{{\color{black}#1}}

\begin{document}
\title{Low-Complexity Soft-Decision Detection for Combating DFE Burst Errors in IM/DD Links}

\author{Kaiquan~Wu,~\IEEEmembership{Graduate Student Member, IEEE}, Gabriele~Liga,~\IEEEmembership{Member, IEEE}, Jamal~Riani,~\IEEEmembership{Member, IEEE}, and Alex~Alvarado,~\IEEEmembership{Senior Member, IEEE}
\thanks{This work is supported by the Netherlands Organisation for Scientific Research via the VIDI Grant ICONIC (project No. 15685). The work of Alex Alvarado is supported by the European Research Council (ERC) under the European Union’s Horizon 2020 research and innovation programme (grant agreement No. 757791).}
\thanks{Kaiquan~Wu, Gabriele~Liga, and Alex~Alvarado are with the Information and Communication Theory Lab, Signal Processing Systems Group, Department of Electrical Engineering, Eindhoven University of Technology, Eindhoven 5600 MB, The Netherlands (e-mails: \{k.wu, g.liga, a.alvarado\}@tue.nl).}
\thanks{Jamal~Riani is with Marvell Technology, Santa Clara, CA, United States (e-mail:  rjamal@marvell.com).}}


\markboth{Preprint, \today}%
{Shell \MakeLowercase{\textit{et al.}}: Bare Demo of IEEEtran.cls for IEEE Journals}

\maketitle

\begin{abstract}
The deployment of non-binary pulse amplitude modulation (PAM) and soft decision (SD)-forward error correction (FEC) in future intensity-modulation (IM)/direct-detection (DD) links is inevitable. However, high-speed IM/DD links suffer from inter-symbol interference (ISI) due to bandwidth-limited hardware. Traditional approaches to mitigate the effects of ISI are filters and trellis-based algorithms targeting symbol-wise maximum a posteriori (MAP) detection. The former approach includes decision-feedback equalizer (DFE), and the latter includes Max-Log-MAP (MLM) and soft-output Viterbi algorithm (SOVA). Although DFE is easy to implement, it introduces error propagation (EP). Such burst errors distort the log-likelihood ratios (LLRs) required by SD-FEC, causing performance degradation. On the other hand, MLM and SOVA provide near-optimum performance, but their complexity is very high for high-order PAM. In this paper, we consider a one-tap partial response channel model, which is relevant for high-speed IM/DD links. We propose to combine DFE with either MLM or SOVA in a low-complexity architecture. The key idea is to allow MLM or SOVA to detect only $3$ typical DFE symbol errors, and use the detected error information to generate LLRs in a modified demapper. The proposed structure enables a tradeoff between complexity and performance: (i) the complexity of MLM or SOVA is reduced and (ii) the decoding penalty due to EP is mitigated. Compared to SOVA detection, the proposed scheme can achieve a significant complexity reduction of up to $94\%$ for PAM-$8$ transmission. Simulation and experimental results show that the resulting SNR loss is roughly $0.3\sim0.4$ dB for PAM-$4$, and becomes marginal $0.18$ dB for PAM-$8$.

\end{abstract}


\IEEEpeerreviewmaketitle

\section{Introduction}\label{sec:Intro}

\IEEEPARstart{T}{he} growing demands for optical high-speed data traffic stemming from applications such as media services, and cloud computing, has driven data center interconnect (DCI) networks to move forward towards 800Gb/s and beyond. On top of the rate increases, the requirements of low power consumption, low complexity, low cost, and low latency are also extremely important \cite{zhong2018digital,nagarajan2021low}. Intensity modulation (IM) and direct detection (DD) links use only a single photodiode (PD) to receive optical signals, allowing short-reach systems to be highly cost-effective. However, to support large-bandwidth signal transmission, bandwidth-limited transceiver components used in these systems cause severe inter-symbol interference (ISI). For achieving high speeds and overcoming ISI simultaneously, an effective coded modulation (CM) and equalization scheme is highly desired in IM/DD links.

Spectral-efficient modulation formats and soft decision (SD)-forward error correction (FEC) are two indispensable elements for CM in future IM/DD links. Nonreturn to zero (NRZ) modulation is becoming insufficient to support the growing data rates. To go beyond the $1$ bit/symbol limit of NRZ, high-order $M$-ary amplitude modulation (PAM) such as PAM-$4$ and PAM-$8$ \cite[Sec.~III]{gaudet2016survey} is needed. Recently, an SD concatenated code has been adopted as the optical interconnect FEC in 200 Gb/s PAM-$4$ serial intra DCI applications \cite{marvell2023fec_ieee}. As a consequence of the increased number of symbol levels, the noise sensitivity is worsened. FEC is a key enabler for dealing with the worsened noise sensitivity. Indeed, FEC is expected to have an overhead between $7\%$ and $15$\% for the sake of complexity and power consumption, and provide a net coding gain of around $10$ dB \cite[Sec.~IV-C]{nagarajan2021low}. In pursuit of a higher symbol rate, next-generation DCIs are likely to resort to powerful SD-FEC for the additive white Gaussian noise (AWGN) channel. However, the potential of SD-FEC under a memoryless AWGN assumption is yet to be fully exploited due to the performance penalty caused by strong ISI \cite{schaedler2021recurrent,stojanovic2023dfe}.

Equalization techniques aim at removing the ISI from the received symbols. A decision-feedback equalizer (DFE) has a simple structure and is effective at canceling post-cursor ISI \cite{mahadevan2021hard}. The performance of DFE is limited by a phenomenon known as error propagation (EP). Due to the feedback nature of DFE, one symbol decision error causes erroneous subsequent decisions, resulting in burst errors. The statistics of burst errors are usually analyzed by Markov models that describe the evolution of the DFE decisions. For the purpose of post-FEC bit error rate (BER) estimates, several Markov models have been proposed \cite{narasimha2009impact,mahadevan2021hard,yang2019statistical,kim2021accurate}. The analysis becomes complex when multi-tap DFE \cite{yang2019statistical} or residual ISI \cite{kim2021accurate} is considered.

DFE burst errors degrade the error-correcting performance of FEC decoders. For systems using NRZ and hard decision (HD)-FEC, the decoding penalty due to burst errors can be reduced by simple techniques such as bit interleaving or precoding based on differential and modulo operations \cite{russell1995technique,marvell2023dfe_ieee}. These two techniques have been adopted in the IEEE 802.3ca standard \cite{ieee2020PON}. However, when PAM-$M$ and SD-FEC are used, their effectiveness becomes questionable. For example, the error-correcting performance of SD-FEC is sensitive to the accuracy of the input log-likelihood ratios (LLRs). LLRs are often computed using a memoryless AWGN assumption, which is incorrect in the presence of burst errors \cite{schaedler2021recurrent,mahadevan2021impact}. In addition, interleaving might cause an unacceptably high latency, and precoding prevents traditional demapper from accurate LLR computation due to the modulo operation error \cite[Sec.~III-A]{jana2017pre}, making them less attractive. 

To decrease the impact of burst errors, the so-called weighted decision feedback equalizer (WDFE) was proposed for wireless communications in \cite{palicot2000weighted,palicot2008performance}. WDFE exploits the reliability of the symbol's HD and weighs them in the feedback loop. However, a rigorous mathematical derivation of the WDFE reliability is not available. The values are currently computed via heuristic reliability functions \cite{wettlin2022investigation}. Nevertheless, WDFE has been recently studied for short-reach IM/DD links with HD-FEC in \cite{zhou2021burst,zhang2021c}. Significant performance improvements were reported in experiments. A performance evaluation of WDFE for SD decoding is missing.

Compared to DFE, trellis-based detection algorithms are the most effective approach to overcoming ISI. In general, for PAM-$M$ and a channel with memory $L$, such algorithms perform calculations on a fully-connected trellis constructed on a $M^L$-state machine. Maximum likelihood sequence detection (MLSD), which finds the most probable path on the trellis given the received symbols, can be implemented using the Viterbi algorithm (VA). Since we target IM/DD systems using SD-FEC, a \emph{symbol-wise} soft output is necessary for the subsequent bit-wise LLRs. Hence, MLSD is not applicable, but instead, symbol-wise maximum a posteriori (MAP) detection \cite{bahl1974optimal} is needed. Practical MAP detection algorithms include the Max–Log–MAP (MLM) \cite{robertson1997optimal} and the soft-output Viterbi algorithm (SOVA) \cite{hagenauer1989viterbi,cong1999sova,fossorier1998equivalence}. MLM and SOVA achieve near-optimum performance but at the cost of high complexity due to a large number of states ($M^L$). Various state pruning methods have been proposed for MLSD \cite{eyuboglu1988reduced,visintin2007long,yu2020reduced,zhou2022advanced}.

In this paper, \RevA{to reduce the penalty of \emph{linear} ISI on SD decoding in short-reach IM/DD systems,} we propose to combine DFE with a trellis-based detection algorithm (MLM or SOVA). This paper extends our work in \cite{wu2022dfe} by achieving more accurate computation of LLRs. In the proposed scheme, MLM and SOVA are designed to detect only $3$ decision error states resulting from DFE. Then, a state demapper is proposed to merge the outputs of DFE and the detection algorithm to yield LLRs. The numerical and experimental results show that the proposed scheme has low complexity and mitigates the impact of DFE burst errors on the SD-FEC decoding.

The remainder of the paper is organized as follows. Sec.~\ref{sec:Pre} presents a brief review of trellis-based detection algorithms, followed by DFE and its EP. Then, the finite-state machine for $3$ typical DFE errors is introduced in Sec.~\ref{sec:DFE_State}. MLM and SOVA for the DFE-$3$ state detection and the state demapper are introduced in the same section. The performance of the proposed scheme is investigated in Sec.~\ref{sec:Results}. Finally, conclusions are drawn in Sec.~\ref{sec:Conc}.

\emph{Notation}: In this paper, the probability of a discrete \NoRev{random} variable is denoted by $P(\cdot)$. The probability density function (PDF) is denoted by $p(\cdot)$. \NoRev{The variance of a variable is denoted by $\Var{\cdot}$.} Sequences are denoted by boldface letters, e.g., $\boldsymbol{y}$, and if necessary sub/superscripts are used for the boundary, e.g., $\boldsymbol{y}_{a}^{b}\triangleq[y_{a},y_{a+1},\ldots,y_{b}]$. Given \NoRev{values of} mean $\mu$ and variance $\sigma^2$, Gaussian distribution is $\mathcal{N}(\mu,\sigma^2)$. To denote the log version of the corresponding variable, we use $\tilde{\cdot}\triangleq\log(\cdot)$.  

\section{Preliminaries}\label{sec:Pre}

\subsection{Partial Response Channel}

In this paper, we assume that an ideal feedforward equalizer (FFE) cancels all pre-cursor ISI in the IM/DD link. We concentrate on a single-tap post-cursor ISI because the impact of multiple taps is considered to be negligible, as previously done in \cite{mahadevan2021impact,stojanovic2023dfe}. The resulting effective channel is a one-tap partial response (PR) channel, as depicted in the green frame in Fig.~\ref{fig:system_block_diagram}. At time instant $i$ ($i= 1,2,\ldots,N$), given the transmitted symbol $x_i$, the received symbol $y_i$ is written as
\begin{equation}\label{Eq:PR_Chn}
y_i = x_i + h x_{i-1} + n_i,
\end{equation}
where $h$ indicates the magnitude of the interference, and $n_i$ stands for the AWGN. Given the noise variance $\sigma^2$, we have $n_i\sim\mathcal{N}(0,\sigma^2)$. As we are targeting short-reach IM/DD systems without amplifiers, bipolar PAM-$M$ is considered \cite{che2021does}. The symbol is drawn from a real set $\mcX$ defined as $ \mcX \triangleq \{\pm1\Delta, \pm3\Delta,\ldots,\pm(M-1)\Delta\}$ for every $m=\log_2 M$ bits. The considered PR channel is illustrated in the dashed frame of Fig.~\ref{fig:system_block_diagram}(b), where $D$ represents a delay element. We assume that the information bits are encoded/decoded using a binary SD-FEC code, and the coded bits $ [b_{i,1}, b_{i,2},\ldots,b_{i,m}]$ are mapped into independent and uniformly distributed symbol $x_i\in\mcX$ with the binary reflected Gray labeling.


\subsection{Decision-Feedback Equalizer and Error Propagation}


\begin{figure*}[t!]
\centering
\setkeys{Gin}{width=0.24\textwidth}
\subfloat[DFE and AWGN demapper.]{\includegraphics[width=\linewidth]{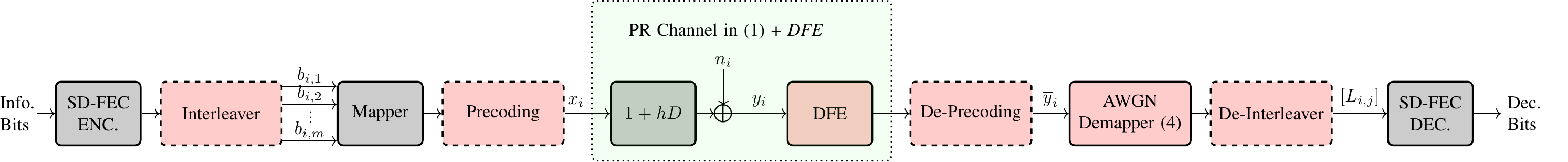}}
\hfill
\subfloat[MLM/SOVA working with PAM-$M$ states.]{\includegraphics[width=0.8\linewidth]{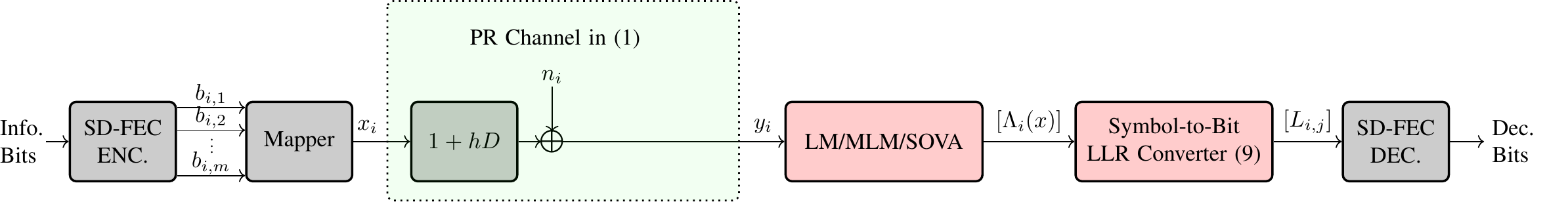}}
\caption{Block diagrams of PAM-$M$ transmission in the one-tap PR channel with different \emph{traditional} detection schemes. In (a), interleaving and precoding (dashed blocks) can be alternatively employed in the DFE-based scheme. }
\label{fig:system_block_diagram}
\end{figure*}


\begin{figure}[t!]
    \centering
    \includegraphics[width=0.6\linewidth]{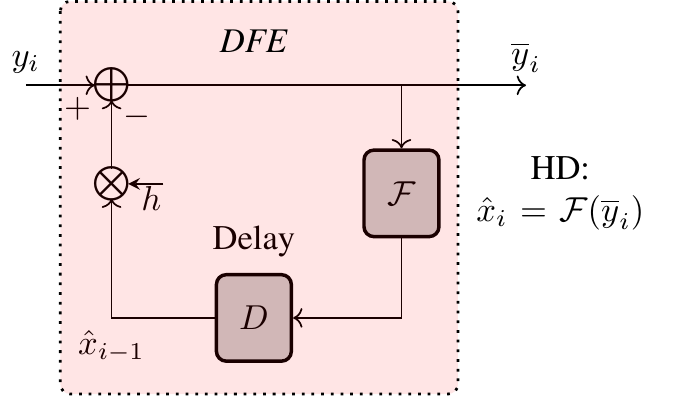}
    \caption{Structure of feedback filter in DFE. The HD at time instant $i$ is used to remove the ISI at instant $i+1$. }
    \label{fig:DFE}
\end{figure}

The feedback structure of DFE for the PR channel transmission is shown in Fig.~\ref{fig:system_block_diagram}(a). In addition to DFE, bit-interleaving or precoding, depicted in dashed blocks in Fig.~\ref{fig:system_block_diagram}(a), can be employed to alleviate the EP. The idea behind DFE is to use the HDs on previous symbols to remove their contribution to ISI from the present received symbol. A one-tap DFE is shown in Fig.~\ref{fig:DFE}. Given the received symbol $y_i$ and the previous HD $\xhat_{i-1}$, the DFE-equalized symbol $\yeq_i $ is computed as

\begin{equation}\label{Eq:DFE_Eq}
\yeq_i = y_i - h \xhat_{i-1}.
\end{equation}
In view of \eqref{Eq:PR_Chn} and \eqref{Eq:DFE_Eq}, the ISI is removed if $\xhat_{i-1}=x_{i-1}$. 

The HD $\xhat_i$ is obtained based on $\yeq_i$ by slicers with thresholds $\{0,\pm2\Delta,\ldots,\pm(M-2)\Delta\}$ located halfway between the symbols from $\mcX$. For example, the HD $\xhat_i$ for PAM-$4$ is

\begin{equation}\label{Eq:DFE_HD}
\xhat_i = \mathcal{F}(\yeq_i)= \left\{\begin{IEEEeqnarraybox}[\relax][c]{l's}
-3\Delta, & for $\yeq_i<-2\Delta$\\
-1\Delta, & for $-2\Delta\leq\yeq_i<0$\\
+1\Delta, & for $0\leq\yeq_i<+2\Delta$\\
+3\Delta, & for $\yeq_i\geq +2\Delta$
\end{IEEEeqnarraybox}\right. .
\end{equation}
The procedures in \eqref{Eq:DFE_Eq}-\eqref{Eq:DFE_HD} repeat to remove ISI from the next received symbol $y_{i+1}$. 

Given the equalized symbols $\byeq=[\yeq_1,\yeq_2,\ldots,\yeq_N]$, a traditional AWGN demapper treats the PR channel and DFE together as an AWGN channel. Therefore, the LLRs are computed as if all ISI has been removed and only AWGN remains. Given the equalized symbol $\yeq_i$, the resulting \emph{mismatched} LLR $L_{i,j}$ is given by
\begin{equation}\label{Eq:AWGN_LLR}
    L_{i,j} = \log \frac{\sum_{x\in\mcX^1_j }p(\yeq_i|x_i=x)  }{\sum_{x\in\mcX^0_j} p(\yeq_i|x_i=x)  },
\end{equation}
where $p(\yeq_i|x_i=x)$ is assumed to follow $\yeq_i \sim \mcN(x,\sigma^2)$, and $\mcX^b_j \subset \mcX$ are the sets of PAM-$M$ symbols labeled by bit $b_{i,j}\in \{0,1\}$ at bit position $j$. Fig.~\ref{fig:system_block_diagram}(a) shows the scheme using DFE and the AWGN demapper to generate bit-wise LLRs for SD-FEC decoding.

\begin{figure}[!t]
    \centering
    \includegraphics[width=1\linewidth]{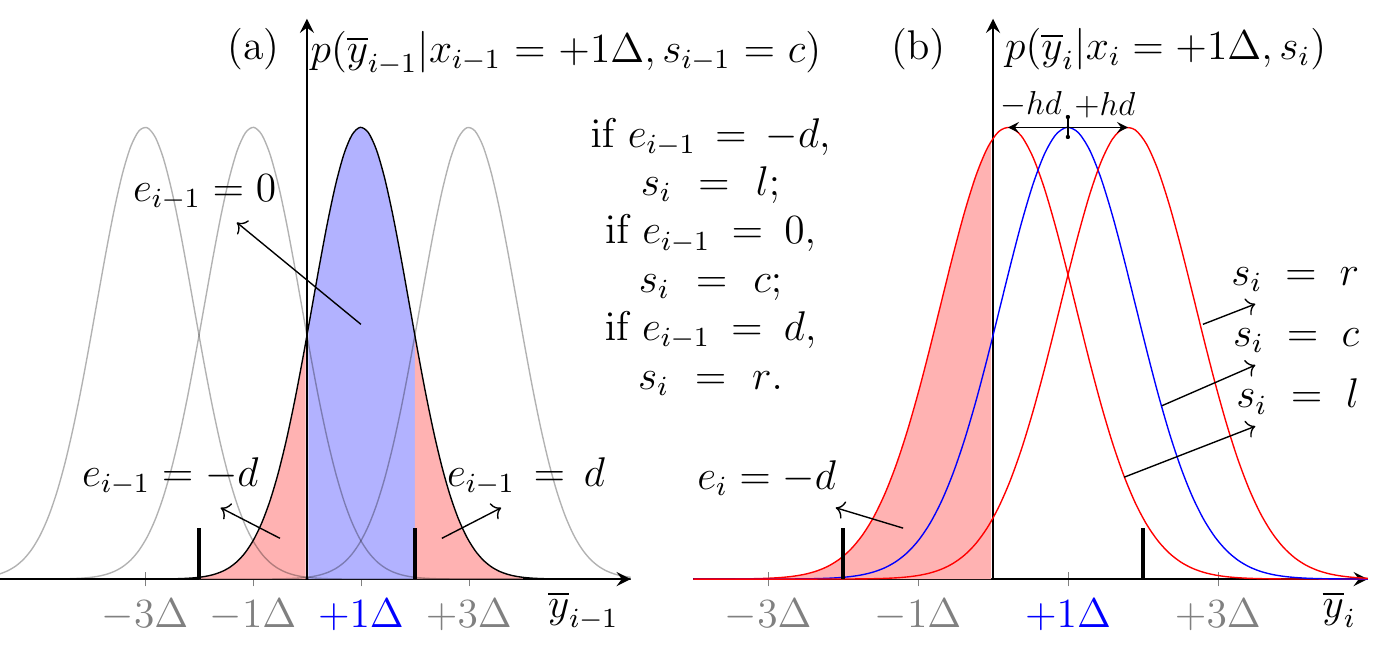}
    \caption{An illustration of the association between the symbol errors $e_{i-1}$ and the biased state $s_i$ via the conditional probabilities $p(\yeq_i|x_i,s_i)$. Here we assume $s_{i-1}=c$ and $x_{i-1}=x_i=+1\Delta$. The colored areas represent $P(e_{i}|x_{i},s_{i})$. The thick vertical lines represent the slicers for decision.}
    \label{fig:pdf_trans}
\end{figure}

The EP of DFE comes from the inherent feedback structure. It can be seen from \eqref{Eq:DFE_Eq} that the removal of ISI for the current equalized symbol $\yeq_i$ depends on a correct prior HD $\xhat_{i-1}$. 
Let the error be defined as $e_{i-1} \triangleq x_{i-1}-\xhat_{i-1}$. For PAM-$M$ with the minimum euclidean distance $d=2\Delta$, there are $(2M-1)$ possible errors, i.e., $e_{i-1}\in\{0,\pm d,\pm 2d,\ldots,\pm(M-1)d\}$. At relatively high signal-to-noise ratios (SNRs) and small $h$, most errors occur between adjacent symbols. Therefore, the most likely errors occurring are $e_{i-1}\in\{0,\pm d\}$ considered instead. After the subtraction in \eqref{Eq:DFE_Eq}, these $3$ errors lead to $3$ different \emph{biased states} $s_{i}$ \NoRev{that are possibly imposed on} $\yeq_i$, which are written as
\begin{equation}\label{Eq:channel_states}
\setlength{\nulldelimiterspace}{0pt}
s_i\triangleq \left\{\begin{IEEEeqnarraybox}[\relax][c]{l's}
 l, & if $e_{i-1}=-d$\\
 c, & if $e_{i-1}=0$\\
 r, & if $e_{i-1}=+d$
\end{IEEEeqnarraybox}\right. .
\end{equation} 
In \eqref{Eq:channel_states}, $s_i\in\mcS\triangleq\{l,c,r\}$ stands for a set of left-biased, center (unbiased), and right-biased states, resp. \RevB{Fig.~\ref{fig:pdf_trans} illustrates the association between the DFE errors and the biased states by displaying the PDF of $p(\yeq_i|x_i,s_i)$. Suppose $s_{i-1}=c$ and $x_{i-1}= +1\Delta$, Fig.~\ref{fig:pdf_trans}(a) shows that the two most relevant symbol errors are $e_{i-1}=-d$ and $d$. When $x_i=+1\Delta$, these errors lead to the left and the right-biased observations of $\yeq_i$ ($s_i=l,r$) resp., as shown in Fig.~\ref{fig:pdf_trans}(b).} From now on, we call $l,c$ and $r$ as ``DFE-$3$ states", and define a function $\omega(s)$ as
\begin{equation}\label{Eq:offset_function}
\setlength{\nulldelimiterspace}{0pt}
\omega(s_i) \triangleq  \left\{\begin{IEEEeqnarraybox}[\relax][c]{l's}
-d, & if $s_i=l$\\
0, & if $s_i=c$\\
+d, & if $s_i=r$
\end{IEEEeqnarraybox}\right. .
\end{equation} 
Using \eqref{Eq:channel_states} and \eqref{Eq:offset_function}, \eqref{Eq:DFE_Eq} can be alternatively written as 
\begin{equation}\label{Eq:yeq_PR}
    \yeq_i=x_i+n_i+h\omega(s_i)=\xhat_i+n_i+h\omega(s_i)+\omega(s_{i+1}).
\end{equation}


In $l$ or $r$ state, the subsequent HD is based on a biased channel observation $\yeq_i$ and thus becomes more prone to an error. \RevB{For the example in Fig.~\ref{fig:pdf_trans}~(b), when $x_i=+1\Delta$ and $s_i = l$, the PDF is shifted to the left by $hd$. Compared to the center state $c$, with the decision threshold $0$, $\xhat_i$ has a higher chance to be $-1\Delta$ at state $l$, as indicated by the areas of $e_{i-1} =-d$ and $ e_i=-d$ in Fig.~\ref{fig:pdf_trans}~(a) and (b).} If error $e_i=-d$ occurs, consequently, the subsequent state is directed to the opposite biased state, i.e., $s_i = l \rightarrow s_{i+1} = r$, which affects again the following decisions. The EP ends only when a correct decision is made in a biased state.

\subsection{\RevA{Weighted DFE}}
\RevA{WDFE mitigates the EP of DFE via an improved decision rule that incorporates the reliability information of HD \cite{palicot2000weighted,palicot2008performance}. Compared to \eqref{Eq:DFE_Eq}, the WDFE-equalized symbol $\yeq_i$ is computed as
\begin{equation}\label{Eq:WDFE_Eq}
\yeq_i = y_i - h \tilde{y}_{i-1},
\end{equation}
where $\tilde{y}_{i-1}$ is a ``hybrid" decision based on the previous HD $\xhat_{i-1}$ and the equalized symbol $\yeq_{i-1}$, i.e.,
\begin{equation}\label{Eq:WDFE_HD}
\tilde{y}_{i-1} = f(\eta_{i-1}) \xhat_{i-1} + \left[1-f(\eta_{i-1})\right]\yeq_{i-1} .
\end{equation}
In \eqref{Eq:WDFE_HD}, $\eta_{i-1}$ is the reliability estimate of $\xhat_{i-1}$, and is used in function $ f(\cdot)$ to produce a weighting factor. The values of $\eta_{i-1}$ and $f(\eta_{i-1})$ range from $0$ to $1$. 

The drawback of WDFE is that the reliability computation is heuristic and thus inaccurate \cite{palicot2008performance}. $\eta_{i}$ is defined as
\begin{equation}\label{Eq:WDFE_eta}
\setlength{\nulldelimiterspace}{0pt}
\eta_{i} \triangleq  \left\{\begin{IEEEeqnarraybox}[\relax][c]{l's} 
1 & if $|\yeq_{i}|\geq(M-1)\Delta$\\
1-\frac{|\xhat_{i}-\yeq_{i}|}{\Delta}, & else
\end{IEEEeqnarraybox}\right. .
\end{equation} 
A popular choice of reliability function $ f(\eta_{i})$ is the sigmoid nonlinear function defined as \cite{palicot2008performance,wettlin2022investigation,zhang2021c}
\begin{equation}\label{Eq:WDFE_f}
f(\eta_{i}) \triangleq  \frac{1}{2} \left(\frac{1-\exp \left[-a(\eta_i/b-1)\right]}{1+\exp \left[-a(\eta_i/b-1)\right] }\right),
\end{equation}
where $a$ ($a>0$) and $b$ ($0\leq b\leq 1$) are user-defined compression parameters. When $b=0$, $ f(\eta_{i})=1$ and \eqref{Eq:WDFE_Eq} is equivalent to \eqref{Eq:DFE_Eq}, and thus WDFE acts as DFE.}

\subsection{Symbol-by-Symbol Soft-Output Detection}

In this section, we give a high-level overview of traditional MLM and SOVA for PR channel detection. Their receiver structure is shown in Fig.~\ref{fig:system_block_diagram}(b). The specific principles of MLM and SOVA will be discussed in detail in Sec.~\ref{sec:DFE_State} (using a different finite state machine model).  

In the presence of ISI, MAP is the optimal rule for symbol-wise detection, in the sense that the a posteriori probability (APP) $P(x_i|\by)$ for each transmitted symbol is maximized. Upon receiving the symbol vector $\by=[y_1,y_2,\ldots,y_N]$, at time instant $i$, the MAP detector evaluates the symbol LLRs $\Lambda_i(x)$ for $x\in\mcX$. Without loss of generality, we consider the symbol $+1\Delta$ as the reference. In this case $\Lambda_i(+1\Delta)=0$, and 
\begin{equation}\label{Eq:Symol_APP}
    \Lambda_i(x) \triangleq \log\left. \frac{P(x_i=x|\by)}{P(x_i=+1\Delta|\by)} \right. , x\in\mcX .
\end{equation}
For the binary SD-FEC decoding, symbol LLRs $\Lambda_i(x)$ must be translated into bit-wise LLRs $L_{i,j}$ ($j=1,2,\ldots,m$). Assuming equally likely bits, such translation is achieved according to
\begin{equation}\label{Eq:Symbol_Bit_LLR}
    L_{i,j} = \log \frac{\sum_{x\in\mcX^1_j } \exp{\Lambda_i(x)}  }{\sum_{x\in\mcX^0_j} \exp{\Lambda_i(x)}  }.
\end{equation}

\begin{figure}[t!]
\centering
\setkeys{Gin}{width=0.24\textwidth}
\subfloat[Finite-state machine of PAM-$4$.
\label{FSM_PAM4}]{\includegraphics[width=0.8\linewidth]{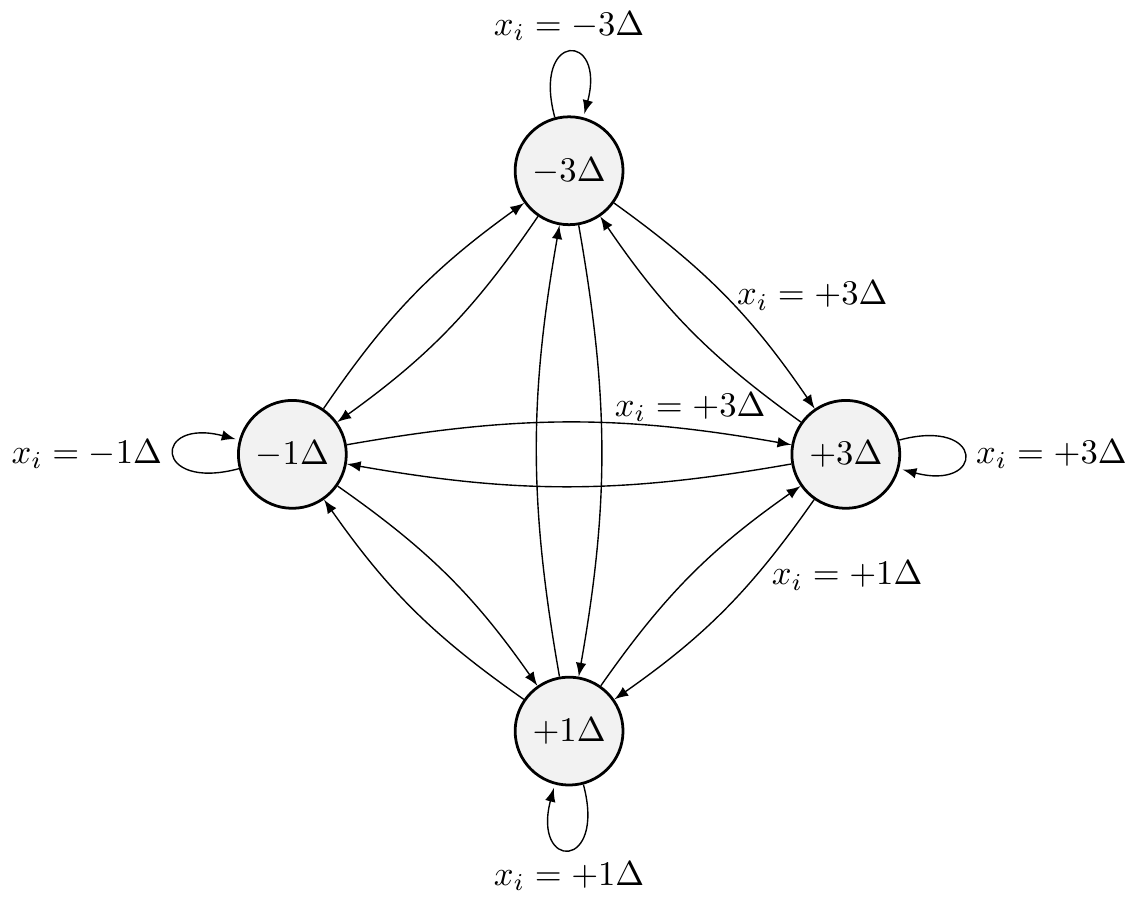}}
\hfill
\subfloat[Trellis of PAM-$4$.
\label{Trellis_PAM42}]{\includegraphics[width=0.8\linewidth]{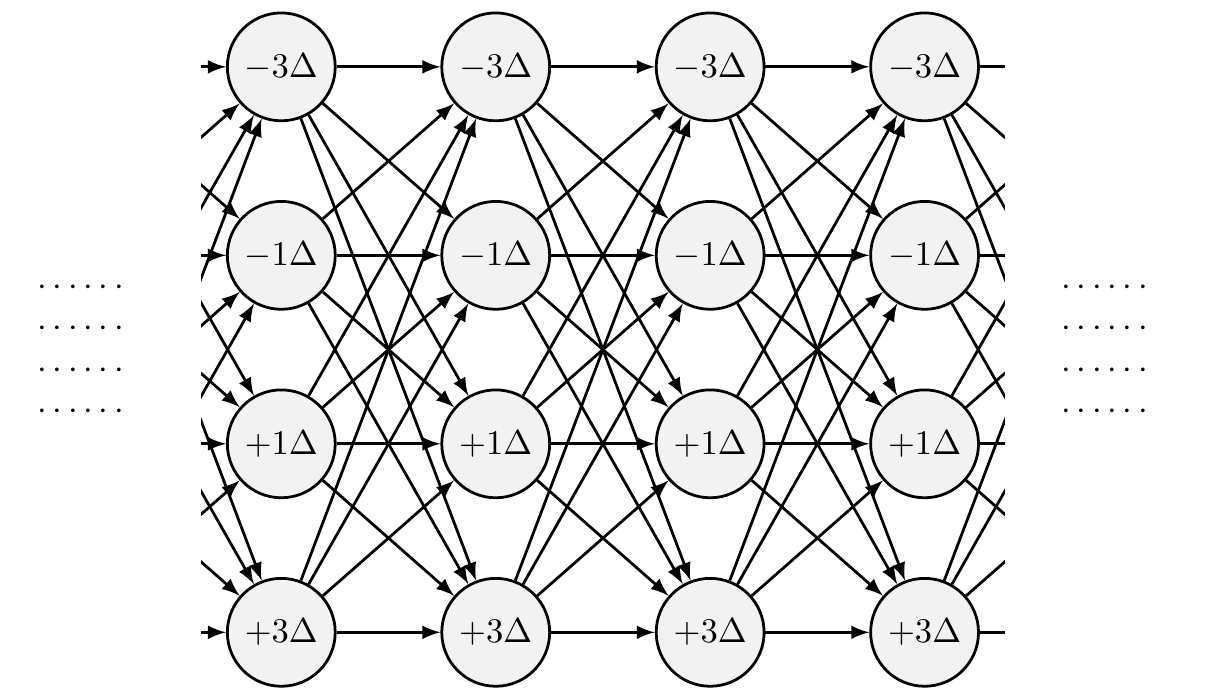}}
\caption{The finite-state machine (a) and the trellis (b) of PAM-$4$ for the one-tap partial response channel in \eqref{Eq:PR_Chn}. In (a), we note that the inputs for only a subset of edges are labeled.} 
\label{fig:PAM4_state_diagram}
\end{figure}

To obtain the symbol LLRs in \eqref{Eq:Symol_APP}, a finite-state machine that characterizes all possible ISI (states), as well as the transitions among them, is used. For the channel in \eqref{Eq:PR_Chn}, the state at time instant $i$ is the transmitted symbol $x_{i-1}$, which contributes to the ISI $h x_{i-1}$ imposed on symbol $x_i$. By unfolding the sates in the time dimension, a fully-connected trellis is obtained, where each path represents a possible transmitted symbol sequence $\bx=[x_1,x_2,\ldots,x_N]$. An illustration of the state machine and the corresponding trellis for PAM-$4$ is shown in Fig.~\ref{fig:PAM4_state_diagram}.
 

There are two popular approaches for obtaining the symbol LLRs in \eqref{Eq:Symol_APP}. One is BCJR to directly compute the APP $P(x_i|\by)$ by performing two recursive computations in the forward and backward directions on the trellis. The implementation of BCJR comes in two flavors, i.e., Log-MAP (LM) and MLM. As will be shown in Sec.~\ref{sec:DFE_State}-C, the difference between LM and MLM is that the APP in the log domain is computed either exactly (in LM), or approximately in (MLM). The other approach is to estimate the difference between two APPs in \eqref{Eq:Symol_APP}, which is used by SOVA. As an augmented version of VA, SOVA performs an additional traceback procedure w.r.t. VA to estimate the symbol LLRs within a \NoRev{length-$\delta$} window. The path metric differences between the surviving path and the nonsurviving paths are the results of their different symbol decisions, which are roughly proportional to the reliability of the decisions. When the traceback window $\delta$ is infinitely long, SOVA is equivalent to MLM, as proved in \cite{fossorier1998equivalence}.


\begin{figure*}
\centering
\includegraphics[width=0.82\linewidth]{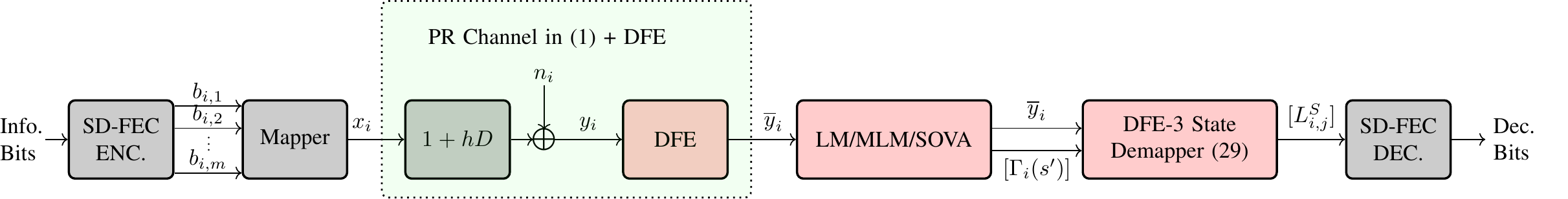}
\caption{Block diagrams of PAM-$M$ transmission in the one-tap PR channel with the \emph{proposed} MLM/SOVA detection working with DFE-3 states.  }
\label{fig:system_block_diagram_DFE3}
\end{figure*}

In terms of implementation, both MLM and SOVA extensively rely on compare and select (CS) operations. Nonetheless, the difference between SOVA and MLM, mainly lies in their backward search on the trellis. SOVA traceback within a limited window, while MLM performs backward recursion \NoRev{on the whole frame of symbols}, making them possess distinct advantages. In general, MLM offers better performance, while SOVA is more friendly for hardware implementation \cite{martin2014further}. For a comparative study between MLM and SOVA, readers are referred to \cite{robertson1995comparison}. The choice between SOVA or MLM often depends on the application under consideration.

\section{Soft-Output Trellis-Based Detection Algorithms of Reduced DFE-3 States}\label{sec:DFE_State}

In this section, \NoRev{we propose a trellis-based soft detection scheme for the PAM-$M$ one-tap PR channels}, whose general structure is shown in Fig.~\ref{fig:system_block_diagram_DFE3}. Instead of \NoRev{producing} the symbol LLRs as expressed in \eqref{Eq:Symol_APP}, the proposed scheme focuses on the LLRs for the $3$ DFE biased states in \eqref{Eq:channel_states}. We first determine the APPs and the LLRs of interests. Then, the corresponding state machine and trellis visualized in Fig.~\ref{fig:state_diagram} will be discussed. \NoRev{Based on this trellis, MLM or SOVA are enabled to detect the DFE-$3$ states.} Finally, a state demapper is proposed to process the output of MLM or SOVA prior to the SD decoder.  

\subsection{Target APP and LLR of DFE-3 States}

The target APP of the biased states for MAP detection is $P(s_i|\byeq)$. This probability can be expressed as

\begin{equation}\label{Eq:APP}
    P(s_i\!=\!s'|\byeq)=\frac{\sum\limits_{s\in\mcS} p(s_i\!=\!s',s_{i+1}\!=\!s,\byeq)}{\sum\limits_{s'\in\mcS} \sum\limits_{s\in\mcS} p(s_i\!=\!s',s_{i+1}\!=\!s,\byeq)}.
\end{equation}For clarity, we denote the joint probability $p_i(s',s,\byeq)\triangleq p(s_i=s',s_{i+1}=s,\byeq)$ and the symbol vector $\byeq_i^{j}\triangleq[\yeq_i,\yeq_{i+1},\ldots,\yeq_{j}]$. Then, $p_i(s',s,\byeq)$ is factored by the chain rule and Markov properties as \cite[Sec.~4.5.3]{ryan2009channel}
 
\begin{equation}\label{Eq:BCJR_Joint_Prob_Factor}
       p_i(s',s,\byeq) =  \alpha_i(s') \gamma_{i+1}(s',s) \beta_{i+1}(s),
\end{equation}
where 
\begin{align}
& \alpha_i(s') \triangleq  p\left(s_i=s',\byeq_1^{i-1}\right), \label{Eq:BCJR_Joint_Prob_Factors_List1} \\
& \gamma_{i+1}(s',s) \triangleq   p(s_{i+1}=s,\yeq_i|s_i=s'), \label{Eq:BCJR_Joint_Prob_Factors_List2}  \\
& \beta_{i+1}(s) \triangleq    p\left(\byeq_{i+1}^{N}|s_{i+1}=s\right). \label{Eq:BCJR_Joint_Prob_Factors_List3} 
\end{align}
The values of $\gamma_{i+1}(s',s)$ in \eqref{Eq:BCJR_Joint_Prob_Factors_List2} are the probabilities that characterize the transition behavior within the DFE-3 states. The values of $\alpha_i(s')$ and $\beta_{i+1}(s)$ are computed through forward and backward recursive inference \cite{bahl1974optimal}. 

The last step (see Fig.~\ref{fig:system_block_diagram_DFE3}) is to calculate the LLRs of the DFE-3 states. These LLRs are denoted by $\Gamma_i(s')$, and defined w.r.t. the center state $c$. Hence, $\Gamma_i(c)=0$ and
\begin{equation}\label{Eq:DFE_LLR}
    \Gamma_i(s') \!\triangleq\! \log \frac{P(s_i\!=\!s'|\byeq)}{P(s_i\!=\!c|\byeq)}\! =\! \log \frac{p(s_i\!=\!s',\byeq)}{p(s_i\!=\!c,\byeq)}, s'\!\in\!\{l,r\},
\end{equation}
where the expression on the right-hand side follows from the fact that $p(s_i\!=\!s'|\byeq)=p(s_i\!=\!s',\byeq)/p(\byeq)$, and we obtain $p(s_i\!=\!s',\byeq)$ by using the total probability law on $p_i(s',s,\byeq)$ in \eqref{Eq:BCJR_Joint_Prob_Factor}, i.e., $p(s_i=s',\byeq)=\sum_{s\in\mcS} p(s_i\!=\!s',s_{i+1}\!=\!s,\byeq)$.

\subsection{DFE-$3$ States and the Transition Probability $\gamma$}




\begin{figure}[t!]
\centering
\setkeys{Gin}{width=0.24\textwidth}
\subfloat[Finite-state machine of DFE-$3$.
\label{FSM_DFE3}]{\includegraphics[width=0.8\linewidth]{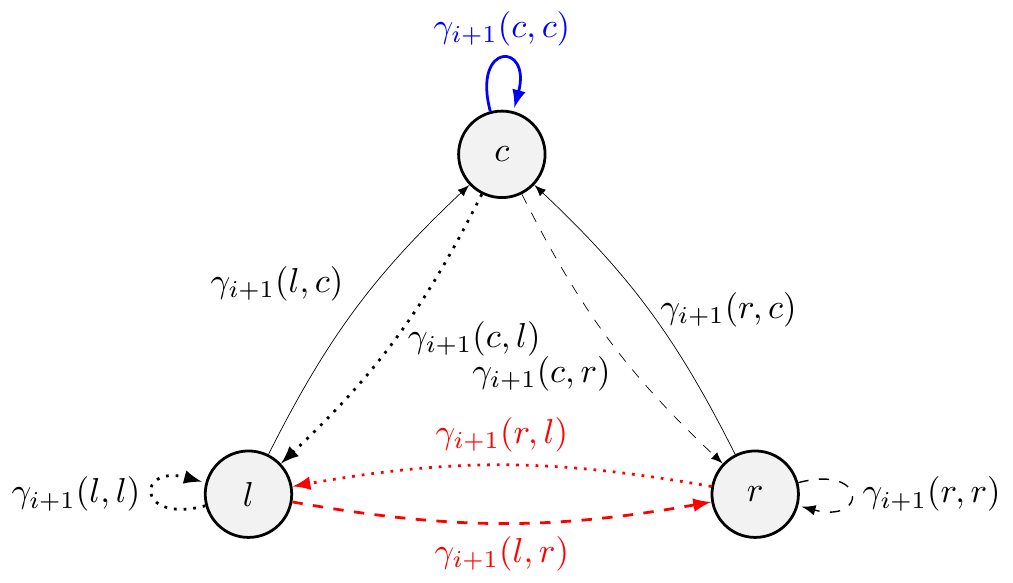}}
\hfill
\subfloat[Trellis of DFE-$3$.
\label{Trellis_DFE3}]{\includegraphics[width=0.9\linewidth]{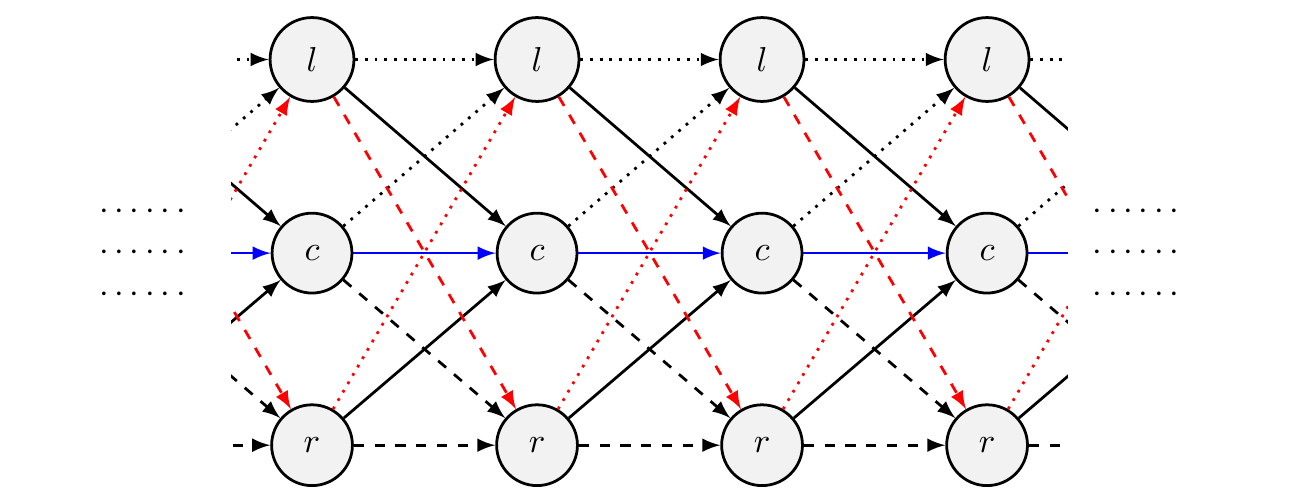}}
\hfill
\subfloat[PAM-$4$ trellis masked by DFE-$3$ trellis.
\label{Trellis_PAM4}]{\includegraphics[width=0.7\linewidth]{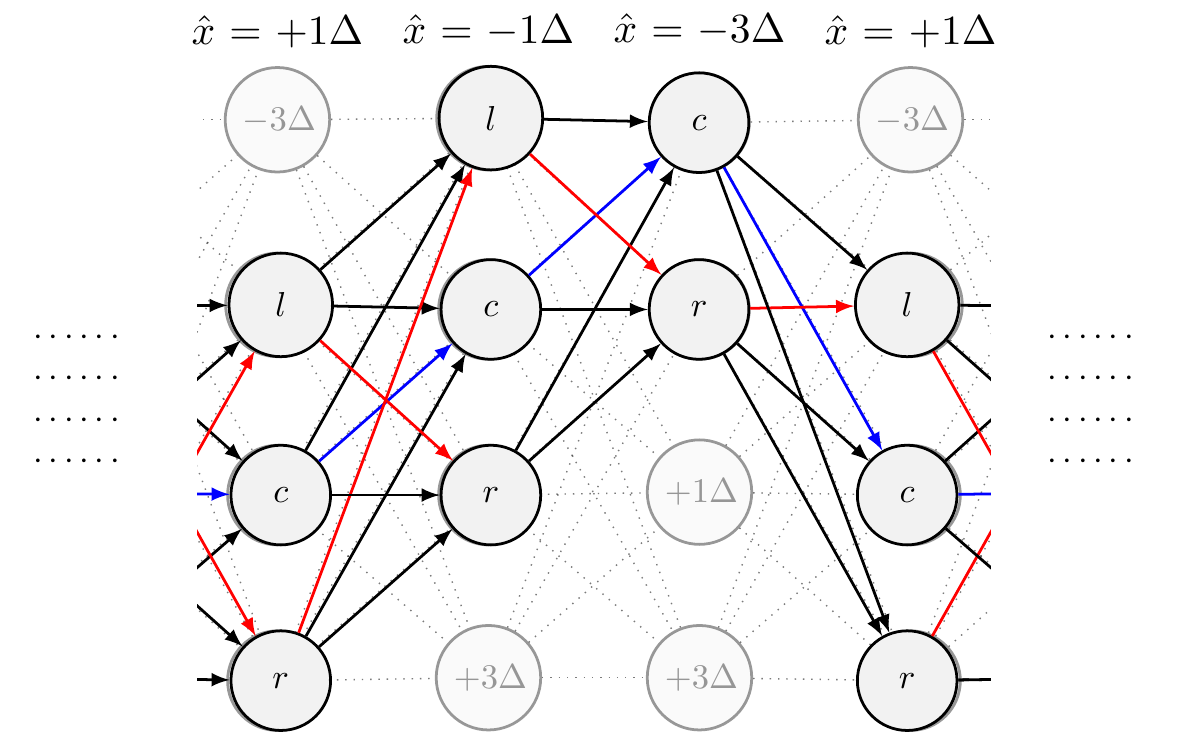}}
\caption{The finite-state machine (a) and trellis (b) of $3$ DFE biased states. (c) shows an example section of DFE-$3$ trellis given certain HD input on top of PAM-$4$ trellis. The blue/red loops indicate error-free and error propagation of the DFE output, resp. The dotted/dashed lines indicate transitions that cannot occur (for the two outermost PAM symbols).}
\label{fig:state_diagram}
\end{figure}


The state machine and the trellis proposed for the DFE-$3$ states are depicted in Fig.~\ref{fig:state_diagram}. For the state machine in Fig.~\ref{fig:state_diagram}(a), each edge labeled with the transition probability $\gamma_{i+1} (s',s)$. There are two major transition loops that characterize the DFE behavior. The transition indicated by the blue loop implies an error-free DFE output, while the EP is represented by the red loop between states $l$ and $r$. As explained in Sec.~\ref{sec:Pre}-B, the self-transitions seldom occur for the biased states, i.e., $\gamma_{i+1} (l,l)$ and $\gamma_{i+1} (r,r)$, and hence are not considered in EP. Fig.~\ref{fig:state_diagram}(b) depicts the corresponding DFE-$3$ trellis. 

Fig.~\ref{fig:state_diagram}(c) compares DFE-$3$ trellis with PAM-$4$ trellis. Given a certain input of HD $\xhat$, the corresponding section of the DFE-$3$ trellis is put on top of the PAM-$4$ trellis in Fig.~\ref{fig:PAM4_state_diagram}(b). The center state $c$ falls upon the HD, and the disconnection to $l$ takes place when $c$ happens to be on the leftmost symbol $-3\Delta$. We conclude that the proposed DFE-$3$ state machine leads to a simpler analysis, because (i) the number of states decreases from $M$ to $3$ and (ii) part of transitions are possibly disconnected and thus leading to a complexity reduction (indicated by the dotted and dashed edges). 

In the following, we explain the computation of $\gamma_{i+1} (s',s)$ with or without disconnections. Given an equalized symbol $\yeq_i$, $\xhat_i=\mathcal{F}(\yeq_i)$ is known, and the next state $s_{i+1}=s$ is determined by $x_i-\xhat_i=\omega(s)$. Hence, $\gamma_{i+1} (s', s)$ in \eqref{Eq:BCJR_Joint_Prob_Factors_List2} is equivalent to
\begin{align}\label{Eq:factorised_gamma}
    & \gamma_{i+1}(s',s) =p(\omega(s),\yeq_i|s_i=s') \nonumber \\
   = & P(x_i=\xhat_i+\omega(s)) p(\yeq_i|s_{i}=s',x_i=\xhat_i+\omega(s)).
\end{align}

When $\xhat_i$ is one of the two outermost symbols, i.e., $\xhat_i \in \{\pm(M-1)\Delta\}$, the transition to either $s=l$ or $r$ is impossible. This is due to the fact that the border symbols have only one neighboring symbol. As a result, one of the HD errors is avoided. 
For example, when $\xhat_i=- (M-1)\Delta$, the transition probability to $l$ is $\gamma_{i+1}(s',l)=0$, because $x_i=\xhat_i+\omega(l)=- (M+1)\Delta\notin\mcX$ causes $P(x_i=\xhat_i+\omega(s))=0$. The probability of ``crossing out" a transition because of this bordering effect is $2/M$ for PAM-$M$. Such disconnection can facilitate end of burst errors and also save the computation for transition probability.

When $\xhat_i$ is an inner symbol, i.e., $\xhat_i\in\{\pm 1\Delta,\pm 3\Delta,\ldots, \pm (M-3)\Delta\}$, $x_i=\xhat_i+\omega(s)\in\mcX$ $\forall s=l$, $c$ and $r$. In this case, $P(x_i=\xhat_i+\omega(s))=1/M$ in \eqref{Eq:factorised_gamma} for equiprobable PAM-$M$. Therefore, we only need to compute $p(\yeq_i|s_{i}=s',x_i=\xhat_i+\omega(s))$. Due to the AWGN $n_i$ in $\yeq_i$ (see \eqref{Eq:yeq_PR}), we have $\yeq_i \sim \mcN(x_i+h\omega(s_i),\sigma^2)$. Hence, in view of \eqref{Eq:factorised_gamma}, $\gamma_{i+1}(s',s)$ is computed as
\begin{equation}\label{Eq:TransProb}
 \gamma_{i\! +\! 1}(s'\! ,\! s) \!\! =\! \! \frac{1}{\sqrt{2\pi}\!\sigma \! M} \! \exp\!\!  \left( \! \frac{\! -\! \left(\yeq_i\! -\! \xhat_i\! -\! h\omega(s')\! -\! \omega(s)\right)^2}{2\sigma^2} \right).
\end{equation}


\begin{figure}[!t]
    \centering
\includegraphics[width=0.65\linewidth]{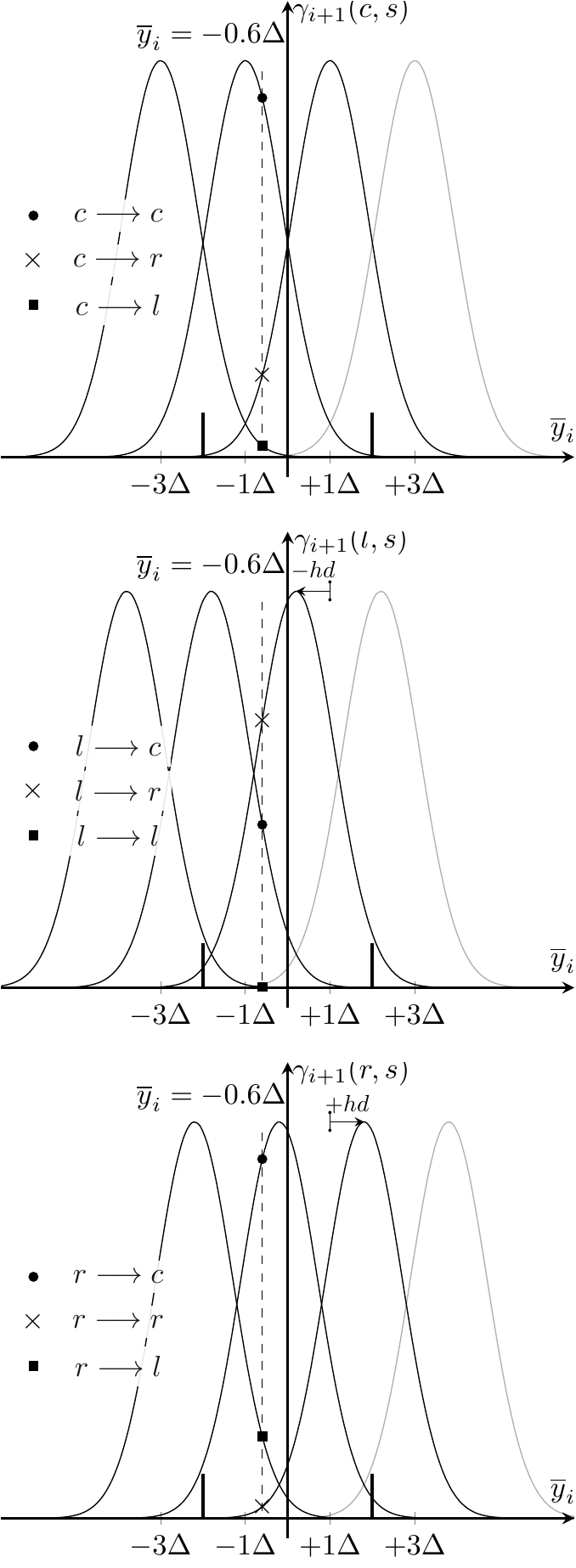}
    \caption{The state transition probabilities $\gamma_{i\! +\! 1}(s'\! ,\! s)$ for $s_i=c$,  $l$, and $r$ are shown in the top, center, and bottom sub-figures, resp. In each sub-figure, suppose $\yeq_{i}=-0.6\Delta$ (implicitly $\xhat_i=-1\Delta$) and $h=0.7$, $\gamma_{i\! +\! 1}(s'\! ,\! s)$ for the transition to states $s_{i+1}=c$, $l$ and $r$ are indicated by the three markers, resp. The thick vertical lines represent the slicers for decision.}
    \label{fig:pdf}
\end{figure}

\begin{example}[$\gamma_{i\! +\! 1}(s'\! ,\! s)$ for PAM-$4$]\label{Example1}
Fig.~\ref{fig:pdf} shows $\gamma_{i\! +\! 1}(s'\! ,\! s)$ for PAM-$4$. Each sub-figure corresponds to a case of state $s_i=c,l,r$ from top to bottom, and consists of $4$ distributions for PAM-$4$ symbols. Specifically, when $\yeq_i=-0.6\Delta $ (indicated by the dashed vertical line), the HD $\xhat_i=\mathcal{F}(\yeq_i)=-1 \Delta$ is an inner symbol, and hence all transitions are possible. The transitions to $s_{i+1}=l,c,r$ states occur when $x_i=-3 \Delta$, $ x_i=-1\Delta$, and $x_i=+1 \Delta$, resp. Therefore, $\gamma_{i\! +\! 1}(s'\! ,\! s)$ are the crossing points between the dashed line and the corresponding distributions for different $x_i$ (depicted in black). 
\end{example}

Typically, $\gamma_{i\! +\! 1}$ in \eqref{Eq:TransProb} is referred to as ``branch metric". In the log domain, the argument of the exponential function in \eqref{Eq:TransProb} denoted as $\lngamma_{i+1}(s',s)$ becomes the branch metric.  
Algorithm~\ref{alg:Branch_Metric} shows how $\lngamma_{i+1}(s',s)$ is computed. Given the DFE equalized symbol $\yeq_i$, $\lngamma_{i+1}(s',s)$ is always computed for transitions to $c$, and possibly skipped for transitions to $l$ and $r$. Note that the factor $1/\sqrt{2\pi}M\sigma$ in \eqref{Eq:TransProb} is omitted, because when eventually computing the DFE-$3$ LLRs as shown in \eqref{Eq:DFE_LLR}, it will appear as a common factor in the denominator/numerator and thus will be canceled out.

\subsection{MLM for DFE-3 States Detection}

\begin{algorithm}[t]
\KwIn{DFE-equalized symbol $\yeq_i$, hard decision $\xhat_i$, noise variance $\sigma^2$, modulation order $M$, biased state set $\mcS$, and tap coefficient $h$ }
\KwOut{ Branch metrics $[\lngamma_{i+1} (s',s)]$, for $s',s \in \mcS$
}

\tcp{Initialize}
$\lngamma_{i+1}(s',s)=-\infty$, $\forall s', s \in{\mcS}$ 



\tcp{For transitions to state $c$}
\For{$s' \in{\mcS}$}{
$\lngamma_{i+1}(s',c) = -(\yeq_i -\xhat_i-h\omega(s'))^2/(2\sigma^2)$
}

\tcp{For transitions to state $l$}
\If{$\xhat_i\neq+(M-1)\Delta$ }{
\For{$s' \in{\mcS}$}{
$\lngamma_{i+1}(s',l) = -(\yeq_i -\xhat_i-h\omega(s')+d)^2/(2\sigma^2)$
}
}

\tcp{For transitions to state $r$}
\If{$\xhat_i\neq-(M-1)\Delta$ }{
\For{$s' \in{\mcS}$}{
$\lngamma_{i+1}(s',r) = -(\yeq_i -\xhat_i-h\omega(s')-d)^2/(2\sigma^2)$
}
}

\KwRet{$[\lngamma_{i+1} (s',s)]$ }
\caption{Branch Metric Computation}
\label{alg:Branch_Metric}
\end{algorithm}

Knowing $\gamma_{i+1}(s',s)$ for $s',s \in \mcS$, $\alpha_{i+1}(s)$ is computed recursively in the forward direction via
\begin{equation} \label{Eq:Fwd_state_Prob_update} 
        \alpha_{i+1} (s)  =   \sum_{s'\in \mcS}  \alpha_i (s')  \gamma_{i+1}(s',s).
\end{equation}
Similarly, the backward recursion is done for $\beta_{i} (s')$ as
\begin{equation} \label{Eq:Bck_state_Prob_update} 
    \beta_{i} (s') =   \sum_{{s}\in\mcS} \gamma_{i+1}(s',s) \beta_{i+1} (s) .
\end{equation}
Without prior knowledge, the values of $\beta_{N+1}(s)$ and $\alpha_{1}(s)$ are initialized as $[\alpha_{1}(l),\alpha_{1}(c),\alpha_{1}(r)]\!=\![\beta_{N\!+\!1}(l),\beta_{N\!+\!1}(c),\beta_{N\!+\!1}(r)]=[0,\!1,\!0]$.

\begin{algorithm}[t]
\KwIn{DFE-equalized symbols $\byeq=[\yeq_1,\yeq_2,...,\yeq_{N}]$, hard decisions $\bxhat=[\xhat_1,\xhat_2,...,\xhat_{N}]$, noise variance $\sigma^2$, modulation order $M$, biased state set $\mcS$, and tap coefficient $h$}
\KwOut{ DFE-3 LLRs $\boldsymbol{\Gamma} = [\Gamma_1(s'),\Gamma_2(s'),...,\Gamma_N(s')]$ for $s'\in\mcS$ }

\tcp{Initialize}
$[\lnalpha_{1}(l),\lnalpha_{1}(c),\lnalpha_{1}(r)]=[-\infty,0,-\infty]$\\ $[\lnbeta_{N+1}(l),\lnbeta_{N+1}(c),\lnbeta_{N+1}(r)]=[-\infty,0,-\infty]$   \\

\tcp{Compute branch metric $\lngamma_{i+1}(s',s)$}
\For{$i=1,2,...,N$}{
\textbf{Algorithm~\ref{alg:Branch_Metric}}
}

\tcp{Forward recursion for $\lnalpha_{i+1}(s)$ in \eqref{Eq:Fwd_state_Prob_Maxlog_update}} 
\For{$i=1,2,...,N$}{
\For{$s \in{\mcS}$}{
$\lnalpha_{i+1}(s) = \max\limits_{s' \in{\mcS}}\left[ \lnalpha_{i}(s') + \lngamma_{i+1}(s',s) \right]$}
}

\tcp{Backward recursion for $\lnbeta_{i}(s')$ in \eqref{Eq:Bck_state_Prob_Maxlog_update}}
\For{$i=N,N-1,...,1$}{
\For{$s' \in{\mcS}$}{
$\lnbeta_{i}(s') = \max\limits_{s \in{\mcS}}\left[ \lnbeta_{i+1}(s) + \lngamma_{i+1}(s',s)\right]$
}

\tcp{Compute $\tilde{p}(s_i=s',\by)$ in \eqref{Eq:APP_log_Factorise}}
\For{$s' \in{\mcS}$}{
$\tilde{p}(s_i=s',\by) = \max\limits_{s \in{\mcS}}\left[ \lnalpha_{i}(s') + \lngamma_{i+1}(s',s) + \lnbeta_{i+1}(s) \right]$}

\tcp{Compute DFE-3 LLRs $\Gamma_i(s')$ in \eqref{Eq:DFE_LLR}}
\For{$s' \in\{l,r\}$}{
$\Gamma_i(s')= \tilde{p}(s_i=s',\by) -  \tilde{p}_i(s_i=c,\by) $
}

}

\KwRet{$\boldsymbol{\Gamma}$}
\caption{DFE-3 MLM}
\label{alg:DFE_MLM}
\end{algorithm}


In practice, \eqref{Eq:Fwd_state_Prob_update} and \eqref{Eq:Bck_state_Prob_update} are implemented in the log domain, which leads to the LM algorithm [Sec.~2.4]\cite{robertson1995comparison}, i.e.,
\begin{equation} \label{Eq:Fwd_state_Prob_log_update}
    \lnalpha_{i+1}(s) = \log\left(\sum_{s'\in\mcS} \exp\left(\lnalpha_{i}(s') + \lngamma_{i+1}(s',s) \right)\right),
\end{equation}
and
\begin{equation} \label{Eq:Bck_state_Prob_log_update}
    \lnbeta_{i}(s') = \log\left(\sum_{s\in\mcS} \exp\left(\lngamma_{i+1}(s',s) + \lnbeta_{i+1}(s) \right)\right).
\end{equation}

Note that the Log-Sum-Exp in \eqref{Eq:Fwd_state_Prob_log_update} and \eqref{Eq:Bck_state_Prob_log_update} can be replaced by the Jacobian logarithm for ease of implementation. Alternatively, one can apply Max-Log (ML) approximation to the Log-Sum-Exp, resulting in MLM. Hence, the recursions are achieved in MLM as
\begin{equation} \label{Eq:Fwd_state_Prob_Maxlog_update}
    \lnalpha_{i+1}(s) = \max\limits_{s' \in{\mcS}}\left[ \lnalpha_{i}(s') + \lngamma_{i+1}(s',s) \right],
\end{equation}
and
\begin{equation} \label{Eq:Bck_state_Prob_Maxlog_update}
    \lnbeta_{i}(s') = \max\limits_{s \in{\mcS}}\left[ \lngamma_{i+1}(s',s) +\lnbeta_{i+1}(s) \right].
\end{equation}
Therefore, in view of \eqref{Eq:BCJR_Joint_Prob_Factor}, MLM computes $\tilde{p}(s_i,\byeq)$ as
\begin{align} \label{Eq:APP_log_Factorise}
     \tilde{p}(s_i=s',\byeq)  =  \max \limits_{s \in{\mcS}}\left[ \lnalpha_{i}(s') + \lngamma_{i+1}(s',s) + \lnbeta_{i+1}(s)   \right].
\end{align}
The final step in DFE-$3$ MLM is to compute the LLRs of biased states $\Gamma_i(s')$ (see Fig.~\ref{fig:system_block_diagram_DFE3}) based on \eqref{Eq:DFE_LLR}. The DFE-$3$ MLM algorithm is summarized in Algorithm~\ref{alg:DFE_MLM}. After initialization, it calls Algorithm~\ref{alg:Branch_Metric} to obtain the branch metrics. Then, $3$ major loops achieve the two recursions and the update of the joint probabilities $\tilde{p}(s_i=s',\byeq)$, resp.

\subsection{SOVA for DFE-$3$ States Detection}


\begin{algorithm}[t]
\KwIn{DFE-equalized symbols $\byeq=[\yeq_1,\yeq_2,...,\yeq_{N}]$, hard decisions $\bxhat=[\xhat_1,\xhat_2,...,\xhat_{N}]$, noise variance $\sigma^2$, modulation order $M$, biased state set $\mcS$, \RevB{traceback length $\delta$}, and tap coefficient $h$}
\KwOut{ DFE-3 LLRs $\boldsymbol{\Gamma} = [\Gamma_1(s'),\Gamma_2(s'),...,\Gamma_N(s')]$ for $s'\in\mcS$  }

\tcp{Initialize}
$[\lnxi_{1}(l),\lnxi_{1}(c),\lnxi_{1}(r)] =[\infty,0,\infty]$  \\

\For{$i=1,2,...,N$}{

\tcp{Compute branch metric $\lngamma_{i+1}(s',s)$}
\textbf{Algorithm~\ref{alg:Branch_Metric}}

\For{$s \in{\mcS}$}{

\tcp{Compute path metric $\lnxi_{i+1}(s)$ in \eqref{Eq:DFE_SOVA_PM}}
$\lnxi_{i+1}(s) = \min\limits_{s'\in\mcS} \left[ \lnxi_{i}(s') - \lngamma_{i+1}(s',s) \right]$ \\

\tcp{Update current $\Gamma^s_i(s')$ in \eqref{Eq:DFE_SOVA_Update} }
\For{$s' \in{\mcS}$}{
$\Delta (s',s) = \lnxi_{i}(s') - \lngamma_{i+1}(s',s) - \lnxi_{i+1}(s) $\\
$\Gamma^s_i(s') = \Delta  (s',s)$
}

\tcp{Traceback past $\Gamma^s_j(s')$ in \eqref{Eq:DFE_SOVA_traceback}}
\For{$j=i-1,i-2,...,i-\delta$}{
\For{$s'' \in{\mcS}$}{
$\Gamma^s_j(s'')  = \min\limits_{s'\in\mcS}[\Gamma^{s'}_j(s'')+ \Delta(s',s) ]$
}
}
}
}

$s^{\star}=\argmin\limits_{s\in\mcS}\left(\lnxi_{N+1}(s)\right)$\\
\For{$i=1,2,...,N$}{
\For{$s' \in\{l,r\}$}{
\tcp{\RevB{Estimate DFE-3 LLRs $\Gamma_{i}(s')$ with the smallest path metric }}
$\Gamma_{i}(s') \approx \Gamma^{s^{\star}}_{i}(s')-\Gamma^{s^{\star}}_{i}(c)$
 }
 }
 
\KwRet{$\boldsymbol{\Gamma}$}

\caption{DFE-3 SOVA}
\label{alg:DFE_SOVA}
\end{algorithm}
 

For the DFE-$3$ trellis in Fig.~\ref{fig:state_diagram}, at each time, every state is connected with the previous states through edges. The path with the smallest path metric is chosen as the surviving path for the present terminating state. Let the path metric of the surviving path that terminates in $s_{i+1}=s$ be denoted as $\lnxi_{i+1}(s)$. SOVA calculates $\lnxi_{i+1}(s)$ in the same way as MLM calculates $\lnalpha_{i+1}(s)$ in \eqref{Eq:Fwd_state_Prob_Maxlog_update}, but with opposite signs for indicating the difference between trellis branches, i.e.,
\begin{equation}\label{Eq:DFE_SOVA_PM}
    \lnxi_{i+1}(s) = \min\limits_{s'\in\mcS} \left[ \lnxi_{i}(s') - \lngamma_{i+1}(s',s) \right].
\end{equation}
Similarly, the initialization of $\lnxi_{1}(s)$ is done as that of $\lnalpha_1(s)$ in Algorithm \ref{alg:DFE_MLM}, except with the opposite sign, i.e., $[\lnalpha_1(l),\lnalpha_1(c),\lnalpha_1(r)]\!=[-\infty,\!0,\!-\infty]$.

Then, for the terminating state $s_{i+1}=s$, SOVA approximates the reliability of state $s_{i}=s'$, i.e. $\Gamma^s_i(s')$, to be the cumulative path metric differences between the competing paths and the current surviving path, i.e.,
\begin{equation}\label{Eq:DFE_SOVA_Update}
    \Gamma^s_i(s') = \Delta (s',s) \triangleq \underbrace{\lnxi_{i}(s') - \lngamma_{i+1}(s',s)}_{\text{competing path}} - \underbrace{\lnxi_{i+1}(s)}_{\text{surviving path}}  .
\end{equation}
It can be seen that $\Gamma^s_i(s')$ consists of $3$ sets of reliability of $s_i=s'$, each accounts for a terminating state $s_{i+1}=s$.

The past reliabilities were computed based on the metric differences from the previous surviving path, which is not necessarily the final surviving path. Since the present surviving path has been extended, these past DFE-3 LLRs need an update. SOVA looks back to update a limited range of the recent DFE-3 LLRs. The update is done via \cite[Eq. (6)]{cong1999sova}
\begin{equation}\label{Eq:DFE_SOVA_traceback}
    \Gamma^s_j(s'')  = \min\limits_{s'\in\mcS}[\Gamma^{s'}_j(s'')+ \Delta(s',s) ] .
\end{equation}

Eventually, when reaching the end of the trellis, we have $3$ surviving paths for each terminating state. The final surviving path is the one giving the smallest path metric, whose terminating state is denoted as $s^{\star}$. Then, the DFE-$3$ LLRs $\Gamma_i(s')$ are determined based on the $\Gamma^{s^{\star}}_i(s') $ belonging to the final surviving path. The DFE-$3$ SOVA is given in Algorithm~\ref{alg:DFE_SOVA}. 
Note that Algorithm~\ref{alg:DFE_SOVA} is written such that it outputs the DFE-3 state LLRs for the whole frame. However, to achieve low latency, it is possible to output LLRs continuously once they are out of the traceback window.

\subsection{DFE State Soft Demapping}\label{sec:State_Demapper}


The proposed state demapper\footnote{The idea of the state demapper originates from our previous work in \cite{wu2022dfe}. However, the state demapper in \cite{wu2022dfe} uses $P(s_i=s|\byeq_1^{i-1})$ instead of $P(s_i=s|\byeq)$. $P(s_i=s|\byeq_1^{i-1})$ is obtained by normalizing $\alpha_i(s')$ in \eqref{Eq:BCJR_Joint_Prob_Factors_List1}.} computes bit-wise LLRs $L_{i,j}$ after MLM or SOVA, as shown in Fig.~\ref{fig:system_block_diagram_DFE3}. The reliabilities of DFE-$3$ states are used as weighting factors to account for the bias imposed on the $\yeq_i$. As a result, the LLR accuracy is improved and thus benefits the SD-FEC decoding. \NoRev{The state demapper \emph{approximates} the bit-wise LLR in \eqref{Eq:Symbol_Bit_LLR} with the DFE-3 state information, i.e.,}
\begin{align}
     L^{S}_{i,j} \! \triangleq  \! \log \frac{\sum\limits_{s'\in\mcS }\sum\limits_{x\in\mcX^1_j } \! p(\yeq_i|s_i\!=\!s'\!,\!x_i\!=\!x) \! P(s_i\!=\!s'|\byeq) }{\sum\limits_{s'\in\mcS }\sum\limits_{x\in\mcX^0_j }p(\yeq_i|s_i\!=\!s'\!,\!x_i\!=\!x)\! P(s_i\!=\!s'|\byeq)  },\label{Eq:State_LLR}
\end{align}
where $p(\yeq_i|s_i,x_i)$ has been discussed in Sec.\ref{sec:DFE_State}-B. In \eqref{Eq:Symbol_Bit_LLR}, the effects of all biased states (HD errors) have been implicitly taken into account in $\Lambda_i(x)$ from the PAM-$M$ trellis. By contrast, in \eqref{Eq:State_LLR}, only the information of $3$ important biased states $P(s_i\!=\!s'|\byeq)$ are used to compensate the effects. 

In practice, given $\yeq_i$ and $\Gamma_i(s')$, the LLRs are calculated with the ML approximation as
\begin{align}\label{Eq:MaxLog_State_LLR}
     L^{S}_{i,j}= &  \max\limits_{s'\in\mcS}  \left[ \Gamma_i(s')\! -  \! \frac{\left.\min\limits_{x\in\mcX_k^1}\left[(\yeq_i\!-\!x\!-\!h \omega(s'))^2\right]\right.}{2\sigma^2}   \right] \nonumber \\
      & \!-\!\max\limits_{s'\in\mcS}  \left[ \Gamma_i(s')\! -  \! \frac{\left.\min\limits_{x\in\mcX_k^0}\left[(\yeq_i\!-\!x\!-\!h \omega(s'))^2\right]\right.}{2\sigma^2}   \right] .
 \end{align}
The derivation of \eqref{Eq:MaxLog_State_LLR} is given in Appendix~\ref{Proof:State_LLR}.

 

\begin{figure}[t!]
\centering
\setkeys{Gin}{width=0.24\textwidth}
\subfloat[PAM-$4$ at SNR of $15$ dB. ]{\includegraphics[width=0.48\linewidth]{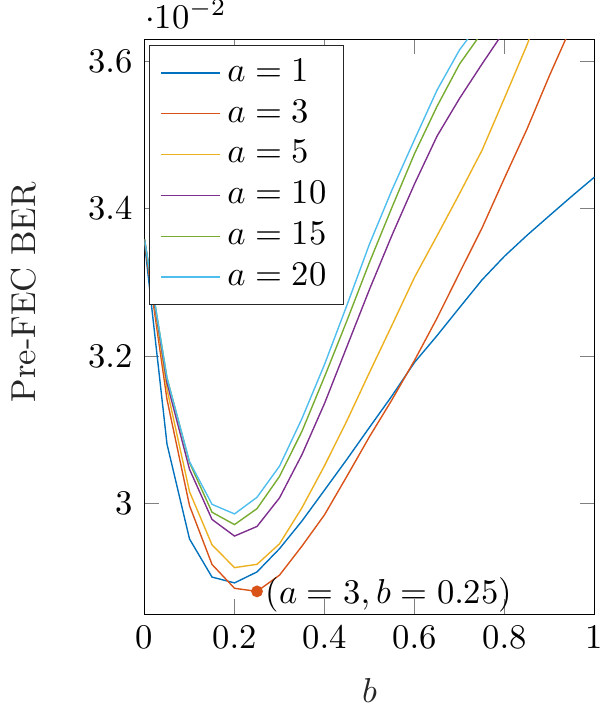}}
\hfill
\subfloat[PAM-$8$ at SNR of $20.5$ dB. ] {\includegraphics[width=0.48\linewidth]{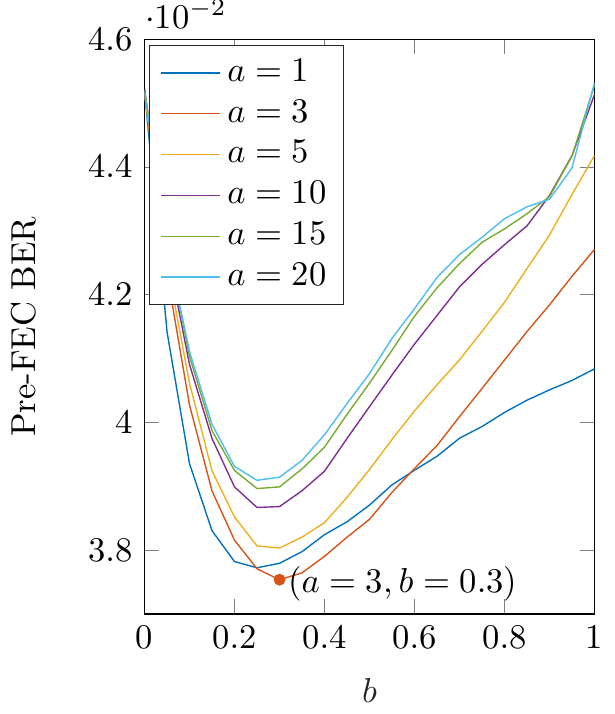}}
\caption{\RevA{Pre-FEC BER vs. the factor $b$ for multiple candidate factors $a$.}}  
\label{Fig:WDFE_ab_optimize}
\end{figure}
 

\section{Results}\label{sec:Results}


\begin{figure*}[t!]
\centering 
\setkeys{Gin}{width=1\textwidth}
\includegraphics[width=1\linewidth]{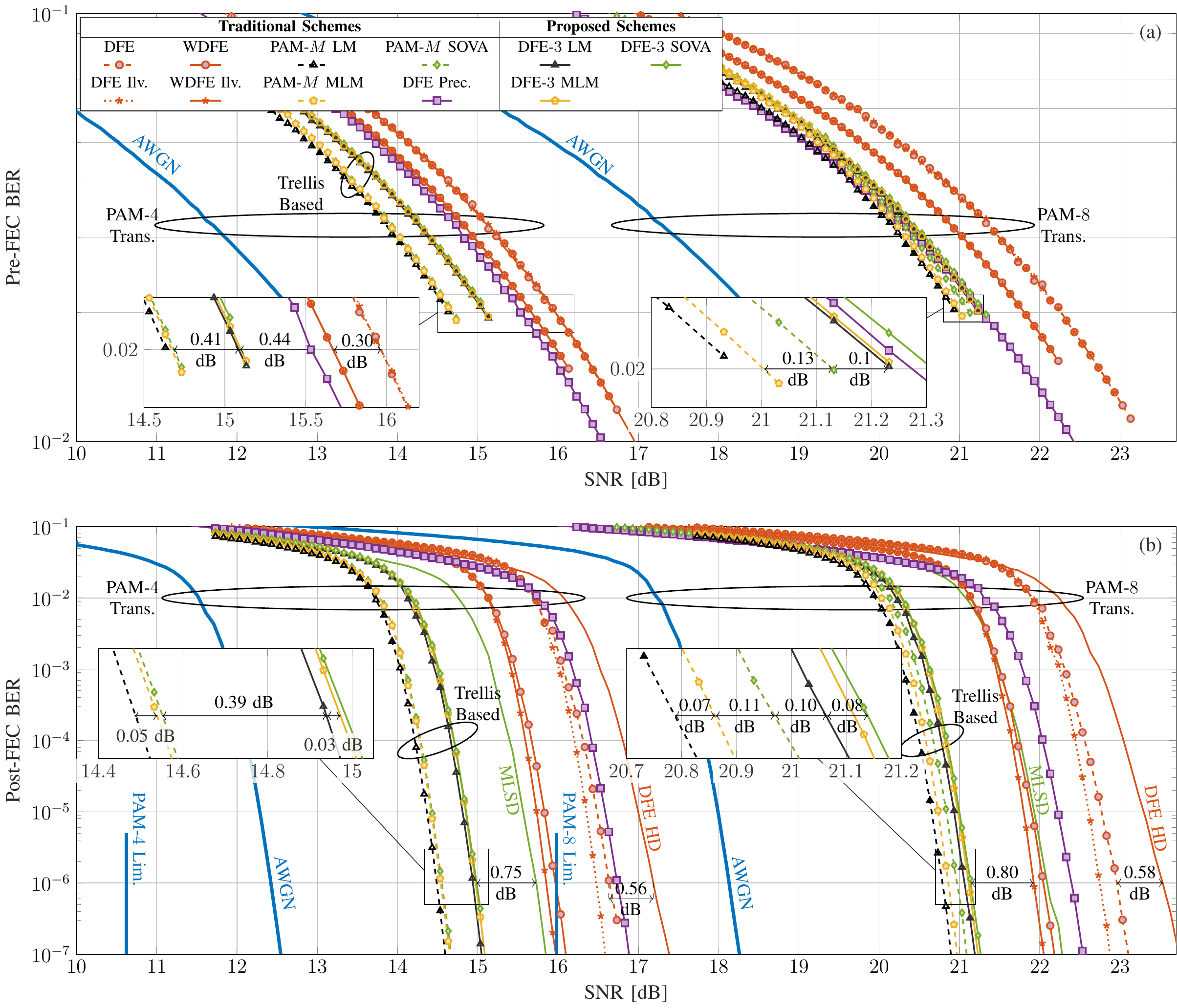}
\vspace{-2em}
\caption{Simulation results for PAM-$4$ and PAM-$8$ transmissions: (a) Pre-FEC BER vs. SNR; (b) Post-FEC BER vs. SNR.}
\label{Fig:Results_Sim}
\end{figure*}

\subsection{Simulation Setup}
\RevC{Based on the effective PR channel estimated in short-reach DCI applications, we consider $h=0.7$ in \eqref{Eq:PR_Chn} to simulate the transmission of PAM-$4$ and PAM-$8$.} The LDPC code from IEEE 802.3ca standard with codeword length of $17664$ and code rate of $R_c\approx0.83$ is employed. The SD decoder performs belief propagation (BP) with $6$ iterations. Transmission in the AWGN channel is simulated and used as a baseline comparison. \RevC{For each considered scheme, at least $500$ codewords are transmitted and at least $10^4$ bit errors are accumulated.} \RevC{To make a fair comparison between the AWGN and the PR channel transmissions, assuming \RevB{unit} signal power, the SNR is defined as $(1+h^2)/\sigma^2$ for the PR channels \cite[Eq. (52)]{forney1972maximum}}, and $1/\sigma^2$ for the AWGN channels (since $h=0$)\footnote{Note that in \cite{wu2022dfe}, we used the AWGN SNR ($1/\sigma^2$) for the PR channel.}.

We focus on the performance comparison between the traditional and the proposed schemes. The proposed schemes in Fig.~\ref{fig:system_block_diagram_DFE3} are referred to as \emph{DFE-$3$ LM/MLM/SOVA}. They require DFE, one of the aforementioned trellis detection algorithms, and the state demapper. The traditional schemes have been depicted previously in Fig.~\ref{fig:system_block_diagram} and include
\begin{itemize}
    \item \emph{DFE}: DFE plus AWGN demapper
    \item \RevA{\emph{WDFE}: WDFE plus AWGN demapper }
    \item \emph{DFE/WDFE Int.}: \emph{DFE} \RevA{or \emph{WDFE}} scheme plus a random bit-interleaver (\RevC{the same interleaver permutation is used for every 4 consecutive codewords}). 
    \item \emph{DFE Prec.}: \emph{DFE} in combination with precoding.
    \item \emph{PAM-$M$ LM/MLM/SOVA}: Trellis-based detection algorithms plus bit-wise demapper.
\end{itemize}
For PAM-$M$/DFE-$3$ SOVA, we set traceback length $\delta=10$. \RevB{In addition, to indicate the SD decoding gains, DFE or MLSD with HD decoding will be presented.} Note that for the HD decoding of precoding, DFE, and MLSD, the demapper uses a lookup table to map each one of the $M$ possible de-precoding outputs into $m$ constant bit-wise LLRs. This table of LLR values is numerically optimized by estimating the crossover probabilities of the equivalent symmetric channel \cite[Ch.~3.4]{szczecinski2015bit}. \RevA{Regarding WDFE, the compression parameters $a$ and $b$ of the sigmoid function in \eqref{Eq:WDFE_f} are optimized from a set of candidate values to yield the lowest pre-FEC BER \cite{zhang2021c}. As shown in Fig.~\ref{Fig:WDFE_ab_optimize}, ($a=3$, $b=0.25$) and ($a=3$, $b=0.3$) are chosen for PAM-$4$/$8$ transmissions, resp. }

\subsection{Simulation Results}

Firstly, the equalizing capability of different schemes is demonstrated in Fig.~\ref{Fig:Results_Sim}(a) via pre-FEC BER vs. SNR performance. The results for PAM-$4$ and PAM-$8$ transmissions are in two clusters. Within the PAM-$4$ cluster, \RevC{non-trellis-based} and trellis-based detection schemes are distinguishable. Among the trellis-based schemes in the left inset, PAM-$4$ LM/MLM/SOVA (dashed lines) correct the most symbol errors, leaving a gap of $0.41$ dB w.r.t. the DFE-$3$ counterparts (solid lines). The DFE-$3$ schemes are followed by DFE with precoding, with a $0.44$ dB penalty. \RevA{Weighting the decision reliability improves the DFE performance by $0.3$ dB, allowing WDFE to approach the performance of DFE with precoding.} Interleaving only breaks the correlation between errors, and thus, the curves for DFE with and without interleaving overlap. 

Comparing the results of PAM-$8$ to PAM-$4$, the following changes are observed: (i) higher SNR is required to achieve the same BER due to increased noise sensitivity. \RevA{(ii) Since PAM-$8$ cause severer EP than PAM-$4$, weighting and precoding improves the pre-FEC BER of DFE more significantly under PAM-$8$ transmission than PAM-$4$, especially for precoding.} (iii) Among the trellis-based schemes, the difference between DFE-$3$ and PAM-$8$ schemes is notably smaller. For example, the right inset shows that PAM-$8$ MLM provides $0.23$ dB gain over DFE-$3$ MLM, while this number for PAM-$4$ is $0.41$ dB. \NoRev{The performance penalty caused by state pruning reduces as the modulation order increases from PAM-$4$ to PAM-$8$}, which has also been previously reported in \cite[Fig.~8]{yu2020reduced}. The intuition behind this reduced penalty is that the performance of the trellis-based schemes deteriorates for an increased number of states due to inherent quasi-catastrophic EP \cite[Sec.~IV-A]{forney1989coset}, and thus, PAM-$8$ suffers from more degradation than DFE-$3$. (iv) SOVA (green curves) is unable to approach the performance of LM and MLM, regardless of PAM-$8$ or DFE-$3$ states. The performance degradation of SOVA implies that using $\delta=10$ is insufficient for PAM-$8$. The reason is that on a large trellis, the paths diverge more easily, and it hampers the accurate update of symbol LLRs within a short window. (v) DFE with precoding outperforms DFE-$3$ SOVA, due to the fact that burst errors become longer and also appear more frequently for PAM-$8$ transmission.

Fig.~\ref{Fig:Results_Sim}(b) shows end-to-end results of post-FEC BER. Large gaps, around $2$ to $3$ dB, are observed between the AWGN capacity (blue vertical lines) and the performances in the AWGN channel (solid blue lines) and the PR channel (other lines). Among the \RevC{non-trellis-based} schemes, \RevA{WDFE exhibits the best performance.} \RevA{For DFE and WDFE, interleaving yields slight performance improvements.} Although precoding provides good pre-FEC BER as has been shown in Fig.~\ref{Fig:Results_Sim}(a), its post-FEC BER is penalized due to the use of LLR lookup table, and this \NoRev{HD} decoding penalty even causes the worst BER results in the PAM-$4$ transmission. \RevC{However, in the presence of severer EP induced by PAM-$8$, such penalty is compensated more by the increased precoding gains in eliminating burst errors.} Nonetheless, the trellis-based schemes can still offer about $1.5$ dB additional gains over precoding. \RevB{Compared to the DFE HD curves, it can be seen taht using SD decoding on DFE brings $0.56$ dB and $0.58$ dB gains for PAM-$4$/$8$, resp.}

\RevA{Furthermore, Fig.~\ref{Fig:Results_Sim}(b) shows that the trellis-based schemes outperform WDFE schemes by more than $0.8$ dB for both PAM-$4$ and PAM-$8$ transmissions.} Comparing the two insets of the trellis-based schemes, we observe that the penalty \NoRev{due to the use of DFE-$3$ states} decreases to around $0.3$ dB for the PAM-$8$ transmission. Especially for PAM-$8$ SOVA (green lines), the right inset shows that the SNR is reduced by $0.18$ dB compared to DFE-$3$ SOVA (at a BER of $10^{-6}$). We also see that for PAM-$8$ transmission, the gap between LM and MLM is $0.07$ dB, which is wider than that for PAM-$4$ transmission, yet still very small. The widened gap is intuitively explained by the fact that the ML used in MLM is more accurate for PAM-$4$ than PAM-$8$ modulation format \cite[Corollary~1]{ivanov2016information}. \RevB{The soft coding gains for VA based algorithms is obtained by comparing the results of MLSD and PAM-$M$ SOVA, which are $1.18$ dB and $0.96$ dB for PAM-$4$/$8$, resp.}
 

\begin{figure}
\centering
\setkeys{Gin}{width=1\textwidth}
\includegraphics[width=1\linewidth]{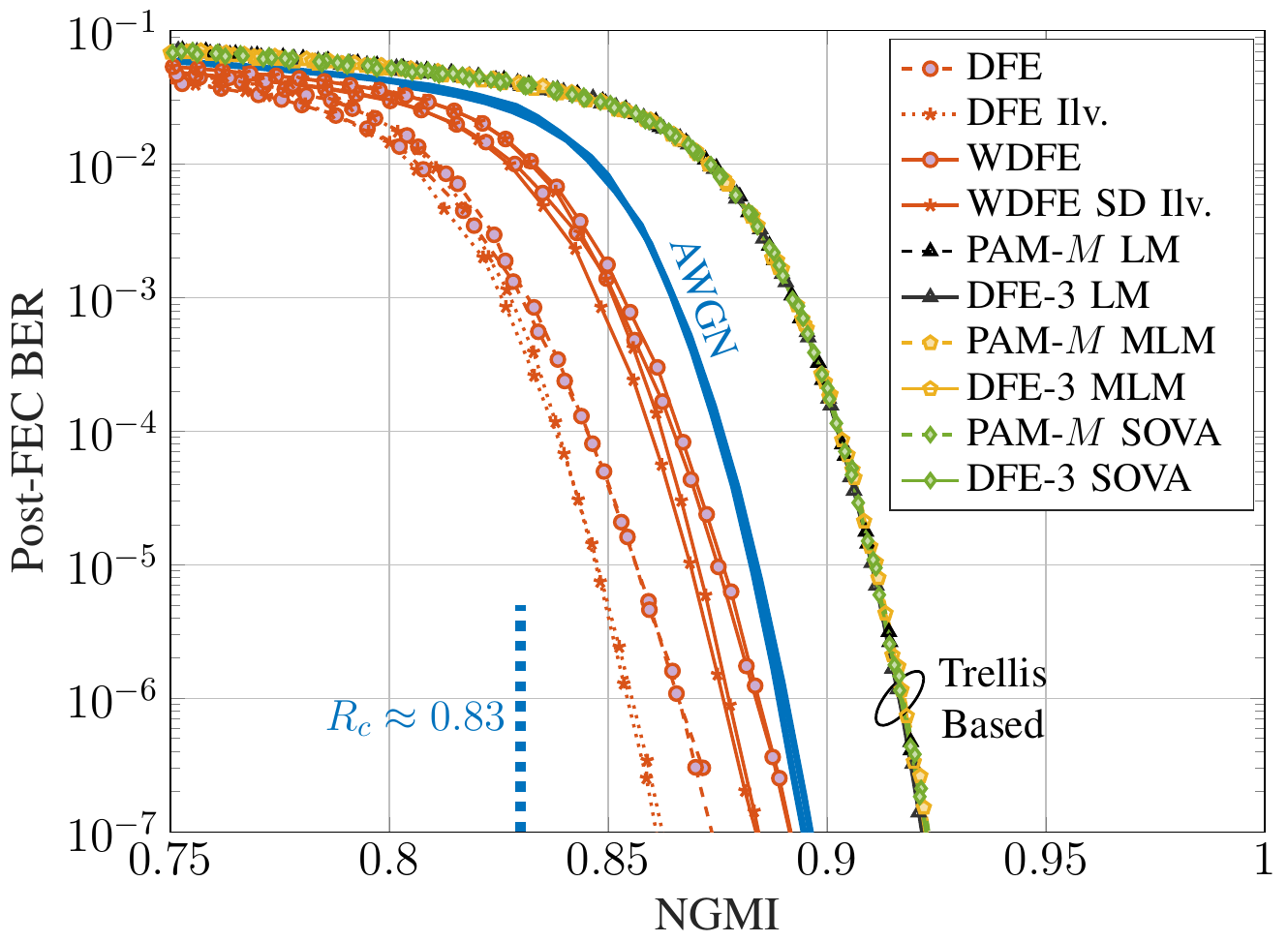}
\caption{ \RevC{Post-FEC BER vs. NGMI of the considered schemes for the simulated PAM-$M$ ($M=4$ or $8$) transmissions.  }}
\label{Fig:Results_Sim_NGMI}
\end{figure}
 



\begin{figure*}[t!]
\centering
\setkeys{Gin}{width=1\textwidth} 
\includegraphics[width=1\textwidth]{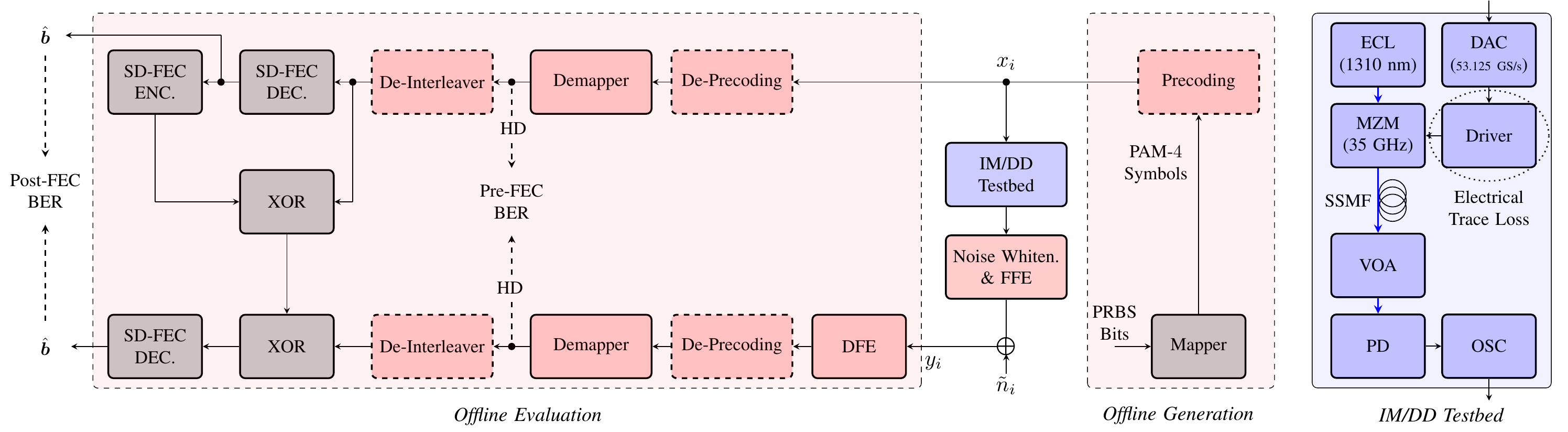}
\caption{Block diagrams of PAM-$4$ IM/DD experiment and offline processing. The dashed pink blocks indicate optional DSPs, depending on the considered schemes. ECL: external cavity laser; DAC: digital-to-analog converter; MZM: Mach-Zehnder modulator; SSMF: standard single-mode fiber; VOA: variable optical attenuator; PD: photo diode; OSC: oscilloscope.}
\label{Fig:Setup_Exp_4PAM}
\end{figure*}

\RevC{Fig.~\ref{Fig:Results_Sim_NGMI} compares the LLR-based normalized generalized mutual information (NGMI) \cite[Sec.~III-F]{alvarado2017achievable} with the post-FEC BER of various schemes. The schemes using HD decoding are excluded. Depending on the considered schemes, the post-FEC BER is achieved at different NGMI than that of the AWGN channel, since LLRs are mismatched to different extents \cite[Sec.~II-B]{alvarado2017achievable}. The modulations PAM-$4$ and PAM-$8$ have no impacts on the results, whose curves for each scheme overlap each other. Specifically, all the trellis-based schemes, achieve the same BER at the almost identical NGMI higher than that of AWGN channel. By contrast, at the same BER, DFE produces LLRs with less NGMI than trellis-based schemes, and similar results have also been reported in \cite{stojanovic2023dfe}. Compared to DFE, WDFE improves the LLR quality and thus increases the NGMI. In addition, applying interleaver to DFE and WDFE does not change the LLR values but breaks their correlations, and thus lower BER is achieved at the same NGMI.}


\subsection{PAM-$4$ Experimental Results}

The performances of the trellis-based schemes have been evaluated in an experiment for PAM-$4$ transmission\footnote{Part of these results were presented in ECOC 2023, Glasgow, Scotland.}. The offline processing and experimental setup are depicted in Fig.~\ref{Fig:Setup_Exp_4PAM}. \RevA{The experiment in Fig.~\ref{Fig:Setup_Exp_4PAM} targets short-reach 100G serial links, operating at a rate of 53.125 GBaud and covering a fiber transmission distance of 500 m.} The memory size of the DSP chip for storing received symbols is 16 Kbytes.~In addition, the testbed suffers from high insertion loss caused by lossy/long electrical traces (e.g., between DAC and driver) and bandwidth limitation.~At the receiver, noise whitening and FFE are performed. \RevA{For the resulting effective channel, the first tap coefficient dominates with a peak magnitude of $0.66$, while the other taps have significantly smaller magnitude $<0.02$. Therefore, the effective channel can be approximated as a one-tap PR channel as in \eqref{Eq:PR_Chn} with $h=0.66$, where the discrepancy mainly lies in the residual ISI due to the use of finite taps.}

In the offline generation, $2^{15}$ bits from a pseudo-random bit sequence (PRBS) are mapped to PAM-$4$ symbols for transmission. The received symbols are stored for offline evaluation. The offline evaluation of the pre/post-FEC BER is based on an emulation approach \cite{schmalen2011generic,stojanovic2013reusing}.~In the upper left corner of Fig.~\ref{Fig:Setup_Exp_4PAM}, scrambling (achieved by XOR operations) and re-encoding are conducted to convert the PRBS-generated codeword into a real codeword. \RevC{For each considered scheme, at least $500$ codewords are transmitted. Given the limited memory size, at most $7525$ codewords can be emulated by applying a sliding window to the stored symbols.} \RevB{Considering only single tap coefficient $h=0.66$, the noise variance after addressing the one-tap ISI is estimated in the electrical domain as $\sigma^2=\Var{y_i-x_i-hx_{i-1}}$. To sweep different SNR values, noise loading is used as shown in Fig.~\ref{Fig:Setup_Exp_4PAM}, where extra offline-generated noise $\tilde{n}_i$ is added.} 

\RevA{Compared to the numerical analysis in the previous section, the traditional schemes under consideration are changed. First, to address the effect of residual ISI, the multi-tap DFE and the EP-free DFE are performed. The EP-free DFE is a genie-aided approach, in which case a DFE always makes correct decisions such that all ISI covered by the feedback loops have been removed \cite{tang2020digital}. Second, the performance of WDFE will not be presented. The reason is that the numerical optimization shows that within our experimental setup, the optimal performance of WDFE is achieved only when it approaches DFE with $b=0$ (see \eqref{Eq:WDFE_f}). This phenomenon has also been reported in \cite{wettlin2022investigation}. }

\begin{figure}
\centering
\setkeys{Gin}{width=1\textwidth}
\includegraphics[width=1\linewidth]{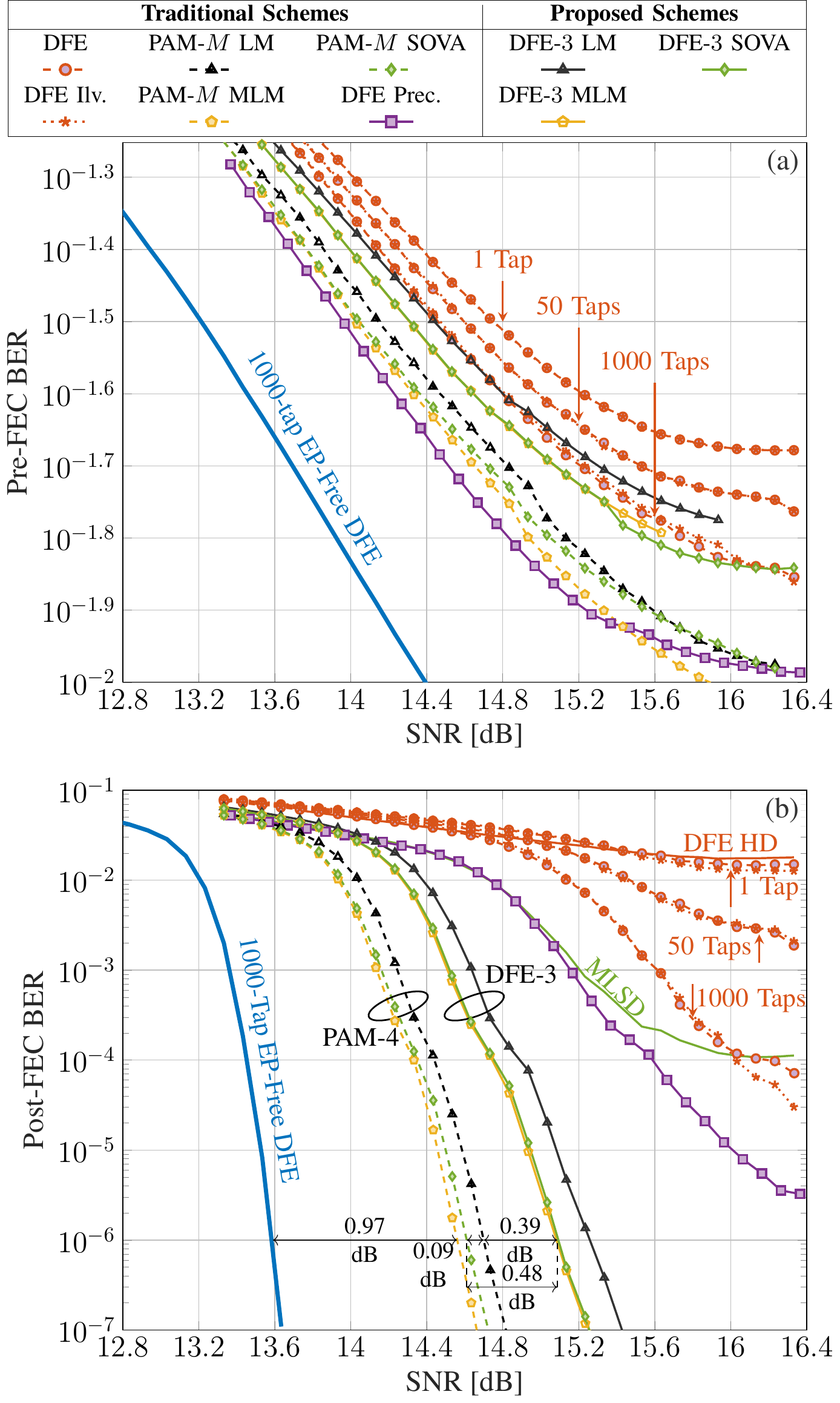} 
\caption{ \RevA{PAM-4 experimental results of: (a) Pre-FEC BER vs. SNR; (b) Post-FEC BER vs. SNR.}}
\label{Fig:Results_Exp_4PAM}
\end{figure}

Fig.~\ref{Fig:Results_Exp_4PAM}(a) shows the pre-FEC BER vs. SNR performance. \RevA{The 1000-tap EP-free DFE exhibits the best pre-FEC BER. However, with the penalty of EP, the performance improvement by increasing the DFE taps becomes marginal \cite{stojanovic2023dfe}. Considerable gains are only observed with a large number of taps, as illustrated by the $50$-tap and the $1000$-tap DFE in Fig.~\ref{Fig:Results_Exp_4PAM}(a).} \RevC{Compared to the simulated PAM-$4$ transmission results in Fig.~\ref{Fig:Results_Sim}, due to the presence of residual ISI, error floors are observed for DFE schemes. The advantage of precoding in correcting burst errors is fully leveraged, and thus DFE with precoding even outperforms trellis-based schemes at SNR lower than $15.2$ dB.} Regarding the trellis-based schemes, using the PAM-$4$ trellis corrects more errors than the DFE-$3$ trellis. Surprisingly, LM exhibits worse performance than MLM and SOVA for both trellises. We conjecture that LM is more sensitive than MLM and SOVA to the presence of residual ISI, because LM follows more strictly the AWGN assumption on the additive noise. The fact that LM is more complex, but ends up with worse performance, makes LM less attractive in real IM/DD systems than MLM and SOVA.



\begin{figure}[t!]
\centering
\setkeys{Gin}{width=1\textwidth}
\includegraphics[width=1\linewidth]{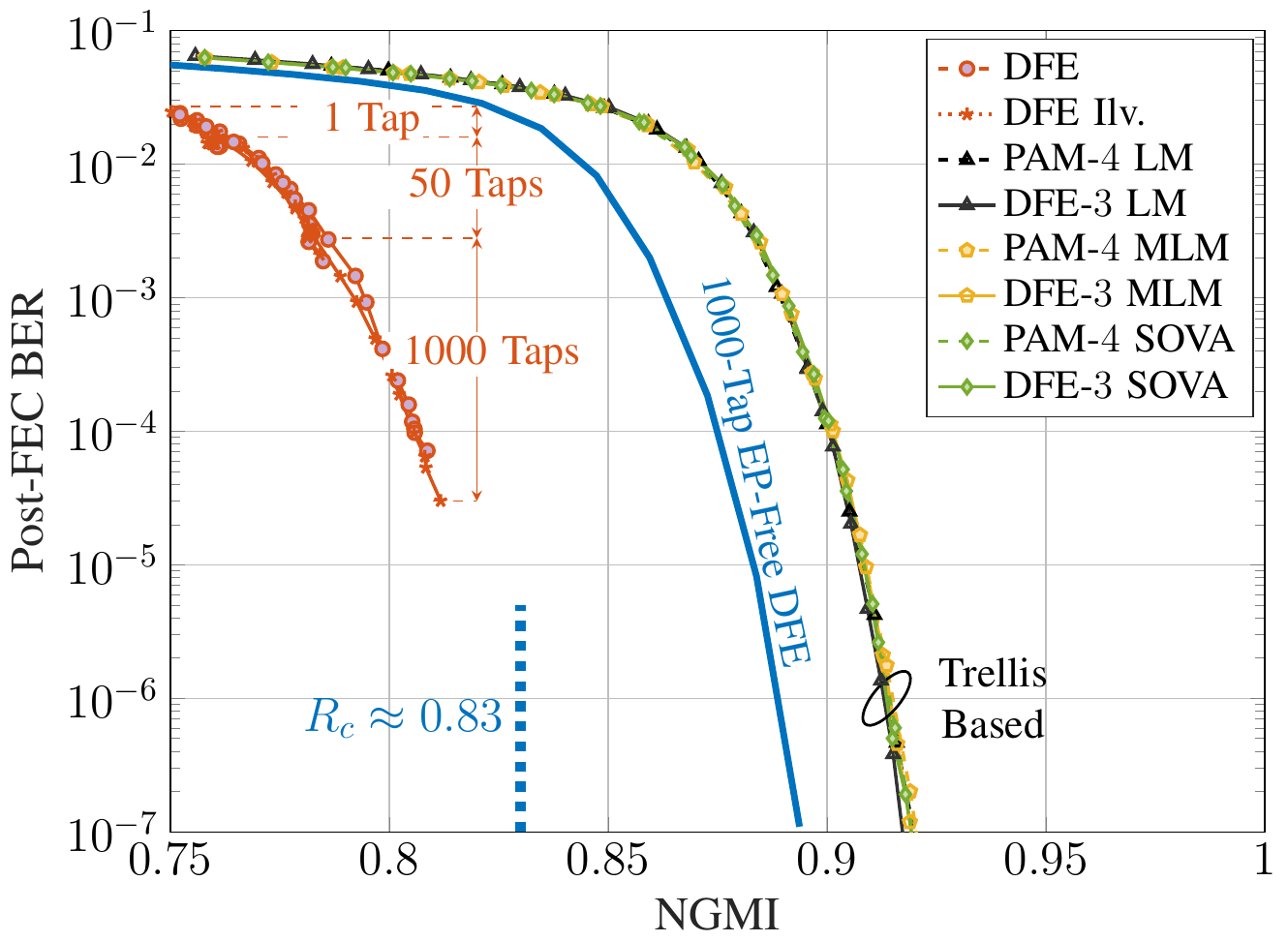} 
\caption{ \RevC{Post-FEC BER vs. NGMI of the considered schemes for the PAM-$4$ experimental results.  }}
\label{Fig:Results_Exp_NGMI}
\end{figure}

Fig.~\ref{Fig:Results_Exp_4PAM}(b) shows the end-to-end results of post-FEC BER. \RevA{The 1000-tap EP-free DFE outperforms other schemes at least by $0.97$ dB. However, with the penalty of EP, the one/multi-tap DFEs exhibit poor performances.} \RevB{Naturally, the one-tap DFE using HD decoding provides the highest BER.} \RevA{Even if the tap number is increased to 1000, DFE cannot reach the BER of $10^{-5}$. In addition, interleaving brings negligible performance improvement.} \RevC{Although DFE with precoding and \RevB{MLSD} yield good pre-FEC BER, as shown in Fig.~\ref{Fig:Results_Exp_4PAM}(a), its post-FEC BER is limited due to the use of discrete LLRs for HD decoding.} On the other hand, Fig.~\ref{Fig:Results_Exp_4PAM}(b) shows that the trellis-based solutions enables better SD decoding performances. \NoRev{Their BER curves are grouped into two clusters, representing the use of PAM-$4$ and the DFE-$3$ states, resp. MLM slightly outperforms LM, and a gain of $0.09$ dB is observed within the PAM-$4$ cluster.} The smallest gap between the DFE-$3$ and the PAM-$4$ clusters is $0.39$ dB. In particular, using the DFE-$3$ states in SOVA leads to a loss of $0.48$ dB.

\RevC{Fig.~\ref{Fig:Results_Exp_NGMI} shows the NGMI vs. post-FEC BER. Compared to Fig.~\ref{Fig:Results_Sim_NGMI}, if we treat EP-free DFE as the case of AWGN channel transmission, it can be seen that they follow approximately the same trends. On the other hand, Fig.~\ref{Fig:Results_Exp_NGMI} shows that DFE schemes exhibit much worse performance in the experiment to such an extent that even employing 1000 taps fails to improve the NGMI to $0.85$.  }

\subsection{Complexity Analysis}

Applying DFE-$3$ states to the trellis-based algorithms offer a big advantage of complexity reduction. In this section, a rough complexity analysis of MLM and SOVA schemes using DFE-$3$ or PAM-$M$ states is presented. We ignore LM in this comparison because MLM is superior to LM in terms of complexity and performance in the experiment, as discussed in the previous section. MLM and SOVA extensively rely on compare and select (CS) operations, making them more computationally intensive than VA which only requires additions and multiplications \cite{robertson1995comparison,rha2019low}. The CS operations correspond to $\max(\cdot)$ and $\min(\cdot)$ in Algorithms \ref{alg:DFE_MLM}-\ref{alg:DFE_SOVA} and \eqref{Eq:MaxLog_State_LLR}. Therefore, this complexity analysis focuses on the number of CS operations. 
For the AWGN/state demapper modules employed in Fig.~\ref{fig:system_block_diagram}(a) and Fig.~\ref{fig:system_block_diagram_DFE3}, and the symbol-to-bit LLR converter in Fig.~\ref{fig:system_block_diagram}(b), ML is used. To make a fair comparison, the CS required inside these modules and DFE are taken into account as well. Furthermore, the number of input elements dominates the complexity of a CS operation and is hence used as a scaling factor for counting. For example, $\max(a_1,a_2,...,a_n)$ is counted as $n$ CS operations.

\begin{figure}[!t]
    \centering
    \includegraphics[width=1\linewidth]{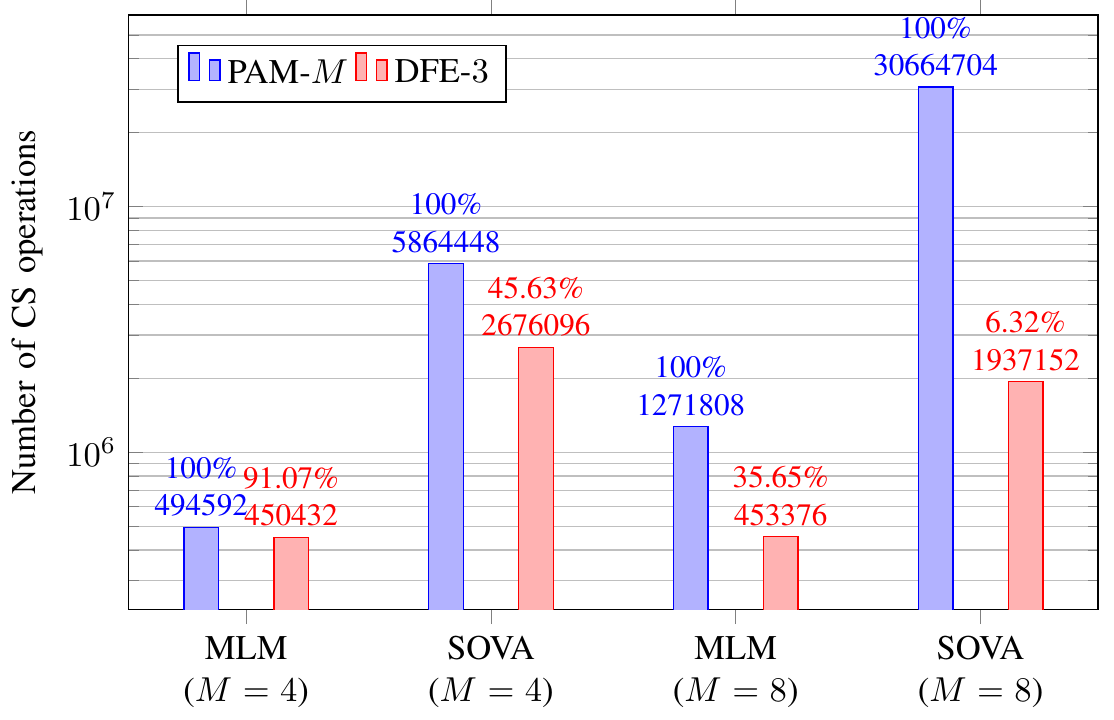}
    \caption{Complexity comparison in terms of compare and select operations between MLM and SOVA ($\delta=10$) when performed on PAM-$M$ or DFE-$3$ state machine. Note that the y-axis is on a log scale.}
    \label{Fig:Ctomplx. Comp.}
\end{figure}

Fig.~\ref{Fig:Ctomplx. Comp.} shows a comparison of the CS operations required per LDPC codeword. \NoRev{It can be seen} that using DFE-$3$ offers considerable complexity reduction for SOVA, mainly because the number of CS is related to the square of the number of states, as can be found in Algorithm~\ref{alg:DFE_SOVA}. 
By contrast, the complexity reduction is less notable for MLM, in which case the number of CS grows linearly with the number of states as shown in Algorithm~\ref{alg:DFE_MLM}. Therefore, the complexity reduction, achieved by decreasing the number of states from $M$ to $3$, is more pronounced, especially for a higher-order $M$. Specifically, the number of CS operations required for the DFE-$3$ SOVA is only $6.32\%$ of the number of CS operations needed for the PAM-$8$ SOVA.



\section{Conclusions}\label{sec:Conc}
In this paper, to improve the decoding of SD-FEC in one-tap PR channels, a scheme incorporating DFE and the symbol-wise soft-output detection algorithm was proposed. We showed that thanks to the pre-processing of DFE, MLM or SOVA only needs to track a trellis with $3$ states associated with a reduced set of DFE symbol errors. The soft information of the symbol errors and the equalized symbols should then be combined to generate bit-wise LLRs. Since the number of states decreases from the number of constellation points to $3$, the complexity of MLM or SOVA is reduced. In particular, the number of compare and select operations for SOVA is reduced by $54\%$ and $94\%$ for PAM-$4$ and PAM-$8$, resp. The complexity reduction comes at the cost of some performance degradation because only a reduced-state trellis is used for the DFE output. Nonetheless, the numerical results of SOVA show that the proposed scheme causes an SNR loss of $0.42$ dB for PAM-$4$ transmission and only $0.18$ dB for PAM-$8$ transmission, \NoRev{and the PAM-$4$ experimental results show that the loss increases slightly to $0.48$ dB}. Overall, the scheme is believed to reach a good trade-off between performance and complexity, which is of interest to future high-speed IM/DD links.

\RevC{Furthermore, an important implication can be drawn from the experimental results: LM is less favored for real-world applications with a large amount of residual ISI, due to its increased complexity and inferior performance compared to its simplified counterpart, MLM.} 

\RevA{Although the proposed scheme specifically targets the one-tap PR channels in this paper, it can be extended to cope with multiple taps. In the general case of PAM-$M$ transmissions in multi-tap ($L>1$) PR channels, the main challenge would be to generalize the rule of identifying the most relevant $3^L$ DFE hard decision errors, and associate them with different states out of the $M^L$-state trellis. This research topic is left for our future work.}

\NoRev{Furthermore, there are other research topics that are worth investigating, including}: (i) the performance and complexity optimization of the state demapper, and (ii) the hardware implementation of the proposed scheme. (iii) Combining the proposed schemes with probabilistic shaping using reverse Maxwell-Boltzmann distribution \cite{che2021does}, \NoRev{which enhances the complexity reduction due to more frequent outermost symbols. }

\appendices
\section{Derivation of State LLR in \eqref{Eq:MaxLog_State_LLR}}\label{Proof:State_LLR}
 
In view of \eqref{Eq:DFE_LLR}, \eqref{Eq:factorised_gamma} and \eqref{Eq:TransProb}, $p(\yeq_i|s_i,x_i)$ and $P(s_i |\byeq)$ can be expressed as exponential functions and inserted in \eqref{Eq:State_LLR}, yielding
\begin{equation}\label{Eq:State_LLR_Appdendix_1}
        L^{S}_{i,j} = \log \frac{\sum\limits_{s'\in\mcS,x\in\mcX^1_j  } \exp\left(\frac{-(\yeq_i\!-\!x\!-\!h \omega(s'))^2}{2\sigma^2} + \Gamma_i(s')\right)  }{ \sum\limits_{s'\in\mcS,x\in\mcX^0_j  } \exp\left(\frac{-(\yeq_i\!-\!x\!-\!h \omega(s'))^2}{2\sigma^2} + \Gamma_i(s')\right)},
\end{equation}
By applying the ML approximation to the numerator/denominator in \eqref{Eq:State_LLR_Appdendix_1}, we have 
\begin{align}
L^{S}_{i,j}= & \max\limits_{\substack{ s'\in\mcS\\x\in\mcX^1_j }}  \left[ \frac{-\left(\yeq_i\!-\!x\!-\!h \omega(s')\right)^2}{2\sigma^2} + \Gamma_i(s')\!  \right] \nonumber \\
 & \!-\!\max\limits_{\substack{ s'\in\mcS\\x\in\mcX^0_j }}  \left[ \frac{-\left(\yeq_i\!-\!x\!-\!h \omega(s')\right)^2}{2\sigma^2} + \Gamma_i(s')\!  \right] \label{Eq:State_LLR_Appdendix_2} \\
 = & \max\limits_{ s'\in\mcS} \! \left[ \max\limits_{ x\in\mcX^1_j}  \! \left[ \frac{-\left(\yeq_i\!-\!x\!-\!h \omega(s')\right)^2}{2\sigma^2}\right] + \Gamma_i(s')\!  \right] \nonumber \\
 & - \max\limits_{ s'\in\mcS} \! \left[ \max\limits_{ x\in\mcX^0_j} \! \left[ \frac{-\left(\yeq_i\!-\!x\!-\!h \omega(s')\right)^2}{2\sigma^2}\right] + \Gamma_i(s')\!  \right] \label{Eq:State_LLR_Appdendix_3} .
\end{align}
After lifting the negative sign out of the max functions over $x\in\mcX^b_j$ in \eqref{Eq:State_LLR_Appdendix_3}, \eqref{Eq:MaxLog_State_LLR} is obtained.


\ifCLASSOPTIONcaptionsoff
  \newpage
\fi

\bibliographystyle{IEEEtran}
\bibliography{DFEbib.bib}
\end{document}